\documentclass[aps,twocolumn,showpacs,superscriptaddress]{revtex4}

\usepackage{graphicx}
\usepackage{amsmath}
\usepackage{amssymb}
\usepackage{amsfonts}
\usepackage{color}
\usepackage{ulem}
\usepackage{textcomp} 
\usepackage{mathcomp} 

\newcommand{\bl}{\color{black}}

\begin{document}

\title{Suppression of polarization decoherence for traveling light pulses via bang-bang dynamical decoupling}

\author{M.~Lucamarini}
\affiliation{School of Science and Technology, Physics Division,
University of Camerino, via Madonna delle Carceri, 9, I-62032
Camerino (MC), Italy.}
\affiliation{{\bl CriptoCam s.r.l., via Madonna delle Carceri, 9,
I-62032 Camerino (MC), Italy.}}

\author{G.~Di~Giuseppe}
\affiliation{School of Science and Technology, Physics Division,
University of Camerino, via Madonna delle Carceri, 9, I-62032
Camerino (MC), Italy.}
\affiliation{{\bl CriptoCam s.r.l., via Madonna delle Carceri, 9,
I-62032 Camerino (MC), Italy.}}

\author{S.~Damodarakurup}
\affiliation{School of Science and Technology, Physics Division,
University of Camerino, via Madonna delle Carceri, 9, I-62032
Camerino (MC), Italy.}

\author{D.~Vitali}
\affiliation{School of Science and Technology, Physics Division,
University of Camerino, via Madonna delle Carceri, 9, I-62032
Camerino (MC), Italy.}

\author{P.~Tombesi}
\affiliation{School of Science and Technology, Physics Division,
University of Camerino, via Madonna delle Carceri, 9, I-62032
Camerino (MC), Italy.}
\affiliation{{\bl CriptoCam s.r.l., via Madonna delle Carceri, 9,
I-62032 Camerino (MC), Italy.}}

\date{\today}

\begin{abstract}
In the propagation of optical pulses through dispersive media, the
frequency degree of freedom acts as an effective decohering
environment on the polarization state of the pulse. Here we
discuss the application of open-loop dynamical-decoupling
techniques for suppressing such a polarization decoherence in
one-way communication channels. We describe in detail the
experimental proof of principle of the ``bang-bang'' protection
technique recently applied to flying qubits in {\bl [Damodarakurup
\underline{\textit{et al.}}, Phys. Rev. Lett. \textbf{103},
040502]}. Bang-bang operations are implemented through
appropriately oriented waveplates and dynamical decoupling is
shown to be potentially useful to contrast a generic decoherence
acting on polarization qubits propagating in dispersive media
like, e.g., optical fibers.
\end{abstract}

\pacs{03.67.Pp, 03.65.Yz, 42.25.Ja, 42.50.Ex}

\maketitle

\section{Introduction}
The interest in the storage and manipulation of quantum systems
have led the researchers to design strategies for preserving the
coherence of such systems against the detrimental effects of the
environment~\cite{Zurek2003}. In fact, the uncontrollable degrees
of freedom of the environment can get entangled with the quantum
system and rapidly destroy the relative phase between the
components of a linear superposition state.

\begin{figure}[t]
   \centering
   \includegraphics[width=.425\textwidth]{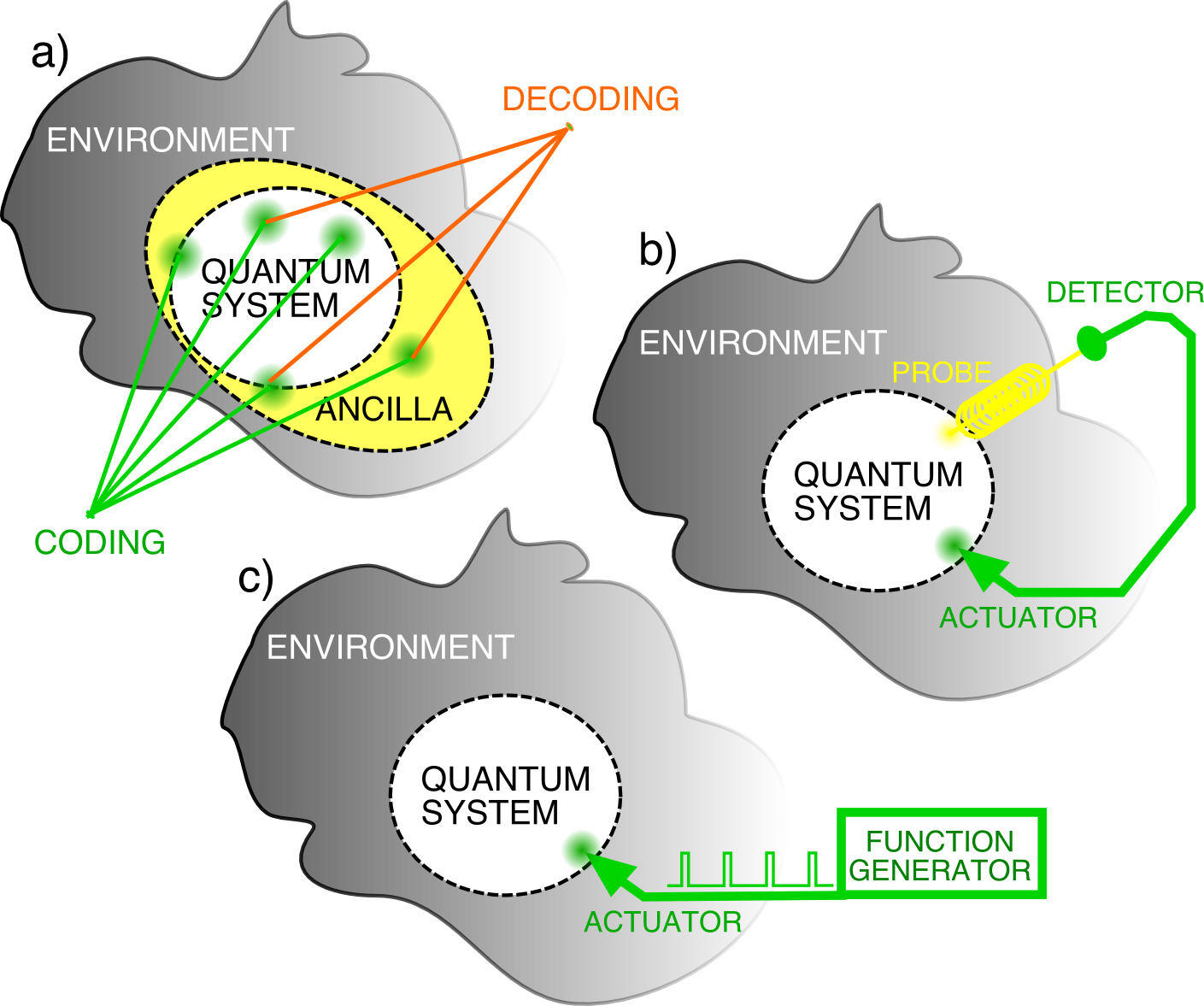}
   \caption{(Color online) Strategies for protecting quantum system against decoherence:
   a) quantum error correction codes and decoherence-free subspaces, both based on state encoding;
    and a number of ancillary systems and on decoding techniques for retrieving the information;
    b) closed-loop techniques, based on measurements and consequent real-time correction of the
    system dynamics; c) open-loop controls, based on external, suitably tailored, time-dependent driving
    chosen such to eliminate undesired effect of the environment.}
   \label{fig:BB_Control_Loops}
\end{figure}

Considerable efforts have been devoted to the issue of
counteracting decoherence. Notable examples are quantum
error-correction codes
(QECC)~\cite{Shor1995,Steane1999,Chiaverini2004,Boulant2005} and
decoherence-free subspaces
(DFS)~\cite{Zanardi1997,Kielpinski2001,Prevedel2007}, both based
on the careful encoding of the quantum state to be protected into
a wider, partially redundant, Hilbert space (see
Fig.~\ref{fig:BB_Control_Loops}). The drawback of these encoding
strategies is the large amount of extra resources
required~\cite{Steane1999}. Alternative approaches which avoid
this hindrance have been developed and may be divided into two
main categories: closed-loop control, also known as quantum
feedback~\cite{Vitali1997}, and open-loop
control~\cite{Viola1998,Viola1999,Vitali1999,Zanardi1999,Kofman2004,Facchi2005}
(see Fig.~\ref{fig:BB_Control_Loops}). In closed-loop techniques,
the system to be protected is subject to appropriate measurements,
whose results are then used for a real-time correction of the
system dynamics. On the contrary, in open-loop controls, the
system is subject to external, suitably tailored, time-dependent
pulses which do not require any measurement. The control
operations are chosen so that any undesired effect of the
environment, such as dissipation, decoherence, heating, can be
eliminated in principle if the controls are applied faster than
the environment correlation time. The physical idea behind these
open-loop schemes comes from refocusing techniques in NMR
spectroscopy~\cite{Ernst1987} where they are typically implemented
via strong and rapid pulses known as ``bang-bang'' (BB) controls.

Open-loop decoupling strategies have been suggested in various
physical contexts: to inhibit the decay of an unstable atomic
state \cite{Agarwal2001a,Kofman2004}, to suppress the magnetic
state decoherence~\cite{Search2000a} and the heating in ion
traps~\cite{Vitali2001}, to contrast the quantum noise due to
scattering processes~\cite{Wu2004} and the polarization mode
dispersion~\cite{Massar2007} in optical fibers. BB control has
been first demonstrated in NMR systems (see
e.g.~\cite{Viola2001}). {\bl Then,} it has been 
{\bl proven with} 
nuclear spin qubits in fullerene~\cite{Morton2006}, 
nuclear-quadrupole qubits~\cite{Fraval2005} and very recently also
with electron spins in Penning traps~\cite{Bollinger2009} and
polarization qubits {\bl in a plane-parallel
cavity~\cite{Berglund2000}~} and in a ring
cavity~\cite{Damodarakurup2009}.

In this paper we discuss the last of the above-mentioned
applications
~\cite{Damodarakurup2009} and provide an analytical explanation of
the results. Photons interact very weakly with their environment
and encoding quantum information in their polarization is a
practical and convenient solution for many quantum communication
protocols~\cite{Gisin2007}. However, the optical properties of
elements like mirrors, crystals and waveplates always depend upon
both polarization and frequency, realizing in this way an
effective coupling between the two degrees of freedom of the
photon. The detection of the final polarization state is then
performed by a detector that inevitably has a nonzero integration
time. Hence the measurement traces out the frequency degree of
freedom and if the field is not monochromatic all the information
transferred to the frequency modes goes lost, inducing on the
polarization qubit  a dephasing process analogous to the
``transversal'' decay occurring in NMR.

Such a polarization decoherence can be inhibited by means of BB
controls implemented in space, i.e., along the photon path, rather
than in time, by means of suitably oriented waveplates. In
Ref.~\cite{Damodarakurup2009} this idea is applied to the
evolution of an optical pulse circling in a triangular ring
cavity. Polarization decoherence occurs because of the reflection
of the pulse on the cavity mirrors, which depend on both frequency
and polarization. After a few cavity round trips, the initially
pure polarization state becomes highly mixed. On the contrary,
after inserting appropriate waveplates in the cavity realizing an
effective impulsive BB, polarization decoherence is significantly
inhibited.

{\bl Previously, a proof-of-principle of the BB impulsive dynamics
was given in~\cite{Berglund2000}, by using optical pulses in a
plane-parallel cavity. That experiment clearly demonstrates that
BB can be applied to light pulses traveling back-and-forth between
two points in space, e.g., on a two-way optical
channel~\cite{Martinelli1989}. On the other hand, the present
experiment including a ring cavity extends the applicability of BB
to one-way channels, because it eliminates the need of a recycling
mirror for suppressing decoherence, and relies entirely on BB
impulsive controls.}

The paper is organized as follows. In Sec.~II we present a brief
introduction on the  dynamical decoupling. In Sec.~III we describe
the experimental apparatus of Ref.~\cite{Damodarakurup2009}, i.e.,
the preparation of the input polarization state, the triangular
ring cavity within which it propagates, and its final detection.
We show how polarization decoherence emerges due to the
polarization dispersive properties of the cavity mirrors. We also
introduce the ``compensated'' cavity obtained by placing an
additional $\mathbb{Z}$ waveplate in the long arm of the cavity;
this is needed if we require that the cavity induces a purely
decohering dynamics on the pulse polarization. In Sec.~IV we
discuss a first experiment in which polarization decoherence is
inhibited through the simple ``Carr-Purcell'' decoupling
scheme~\cite{Viola1998,Viola1999}, which is a particular case of
BB control realized by adding a single wave-plate within the
cavity, suitably oriented. We then consider the more general
Pauli-group decoupling~\cite{Viola1999}, which is realized by
employing \textit{two} waveplates with orthogonal rotation axes
and we show that polarization decoherence is inhibited also in
this case. Finally we modify our ring cavity setup in order to
implement the most general form of decoherence acting on the
polarization qubit. Generic qubit decoherence is realized by
placing a Soleil--Babinet (SB) compensator with axis at
45$^{\circ}$ with respect to the cavity plane in front of each
plane mirror. We show that, in accordance with theoretical
predictions~\cite{Viola1999}, Pauli-group decoupling is effective
in inhibiting decoherence in this general case too. Sec.~V is for
concluding remarks.

\section{The ``bang-bang''--control idea}
The open-loop control technique takes advantage of the finite time
of the relaxation process induced by the environment on the
system. Accordingly, one is able to modify the dynamics by
inducing motions on the system faster than the shortest time scale
accessible by the environment.

For instance, to improve the resolution in NMR spectroscopy,
multiple fast pulse sequences are used in order to reverse the
effect of interactions of spins with the
environment~\cite{Vandersypen2005}. This is obtained through a
careful tailoring of the amplitude, phase, and frequency of the
time-dependent terms in the total Hamiltonian. The main example is
the ``spin echo'' refocussing technique used in NMR, extensively
investigated by E.~L.~Hahn~\cite{Hahn1950}, and later generalized
by Carr and Purcell~\cite{Carr1954}.
In that case, the main sources of noise are the inhomogeneities in
the magnetic field and the molecular spin-spin interactions, which
cause the total magnetization to dephase in a typical relaxation
time called $T_2$. At the microscopic level, the dephasing is due
to the slightly different Larmor frequency experienced by each
spin in the ensemble. In a reference frame which rotates about the
$z$-axis at the Larmor frequency, the $T_2$ process can be
visualized as the spreading of the magnetization vector in the
transverse plane. One can then apply suitable RF pulses at the
Larmor frequency to flip each spin by an angle of $180^{\circ}$.
This realizes an effective ``time reversal'' for the molecules of
the sample and leads, in the ideal case, to the rephasing of the
magnetization vector.

This is analogous to what happens in a standard optical fiber
because of the chromatic dispersion. When a light pulse travels in
the fiber, all its components experience a different local
birefringence, so they propagate at different velocities and the
initial packet broadens. When a second fiber of opposite-sign
dispersion is matched with the first one, a time-reversal of the
frequency components occurs and the chromatic dispersion cancels
out.

A general description of an open quantum system dynamical
decoupling through open-loop control was given by Viola
\textit{\underline{et.~al.}} in Ref.~\cite{Viola99}. Assuming a
quantum system, $S$, described by the Hamiltonian $H_S$, plunged
in an environment described by the Hamiltonian $H_E$, the total
Hamiltonian is given by:
\begin{equation}
    H_{0} =  H_S\otimes \mathbb{I}_E + \mathbb{I}_E\otimes H_E + H_{SE}\,,
\end{equation}
where $H_{SE}$ represents the coupling between system and environment.
The open-system properties like, e.g., the decoherence time, can
be modified by adding to the system Hamiltonian a time-varying
cyclic control, $H_{c}(t)$, with a period $T_C$ such that
\begin{equation}
    U_{c}(t) = \mathcal{T}\exp \left\{-\frac{i}{\hbar}\int_{0}^{t} ds\, H_{c}(s) \right\} = U_{c}(t +
    T_{C})\,.
\end{equation}
In this way the resulting dynamics of the quantum system, usually
called ``stroboscopic dynamics'', is described by an effective
Hamiltonian $H_{eff}$. If the period $T_C$ is sufficiently shorter
than the memory time of the environment, $1/\omega_C$, the
effective Hamiltonian can be represented by the lowest-order
average Hamiltonian~\cite{Viola1998}:
\begin{equation}
    H_{eff} \sim \overline{H_{0}} =\frac{1}{T_C}\int_{0}^{T_C} dt\,
        \left[U_{c}^\dag(t)\,H_{0}\,U_{c}(t)\right]\,.
\end{equation}
Corrections at higher orders in $T_C$ can be systematically
evaluated.

The simplest way of realizing a quantum system dynamical
decoupling is through the quantum BB control, which operates by
repeatedly applying instantaneous, arbitrarily strong, control
pulses (``kicks'') to a qubit, thus disrupting the environmental
interaction. This allows to convert the average over the cycle
into a group-theoreric average, where a discrete set of unitary
control operations, $\mathcal{G} = \{g_0,g_1,\ldots,
g_{|\mathcal{G}| -1}\}$, acting on the quantum system only, are
introduced~\cite{Viola99,Zanardi1999}:
\begin{equation}
    \overline{H_{0}}
    = \frac{1}{|\mathcal{G}|} \sum_{g_j\in |\mathcal{G}|} g_j^\dag\, H_0\, g_j.
\end{equation}
Intuitively the BB symmetrizes the evolution with respect to the
group $\mathcal{G}$, and the system is perfectly decoupled from
the environment when the interaction with the environment is
symmetrized to zero.
\begin{figure}[!ht]
   \centering
   \includegraphics[width=.375\textwidth]{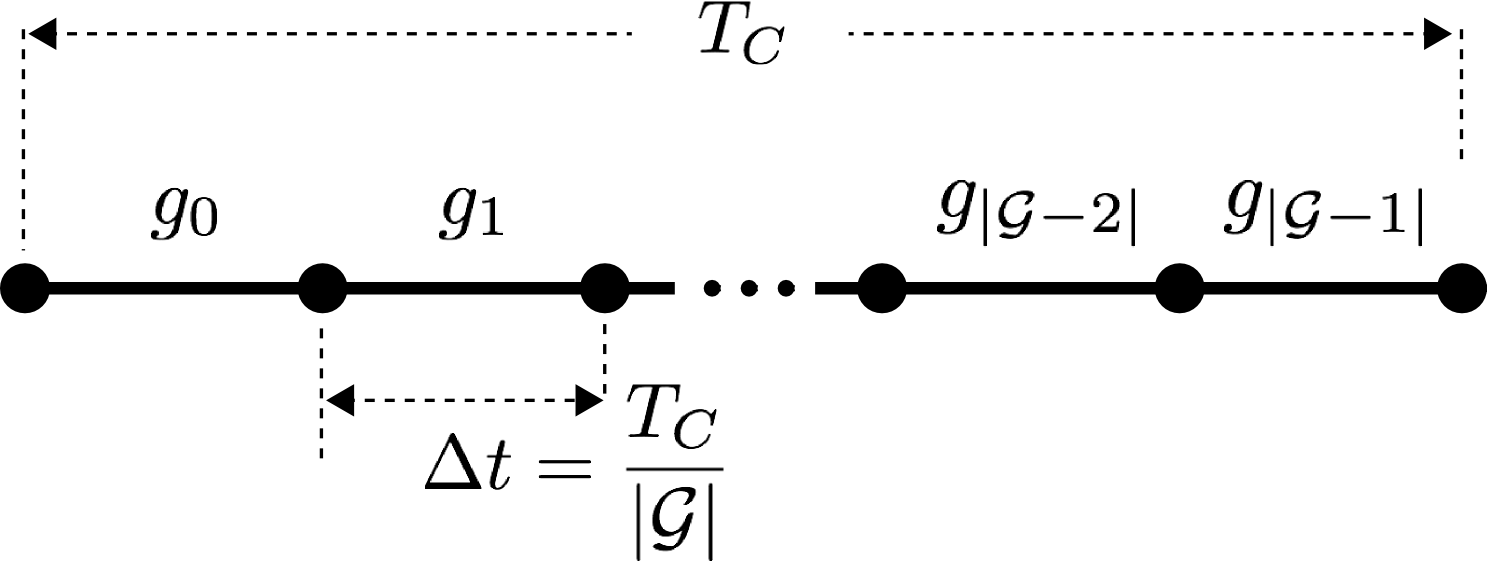}
   \caption{Decoupling group control obtained by the symmetrization under the group $\mathcal{G}$.}
   \label{fig:BB_DecouplingGroup}
\end{figure}

Fig.\ref{fig:BB_DecouplingGroup} provides a graphical
representation of the BB idea. The quantum system enters from the
left side and repeatedly undergoes the periodic transformations
$g_0$--$g_{|\mathcal{G}-1|}$. After each complete cycle in the
group $\mathcal{G}$, the properties of the quantum system are
restored to their initial value (up to corrections at the second
order and higher in $\omega_C T_C$). This allows, in principle, to
indefinitely decouple the system from the environment.

At this point it is straightforward to figure out how the BB
control can be adapted to the optical domain. A light pulse
travels in a free-space portion of space or in an optical fiber
and passes through a series of passive optical elements like
waveplates or phase-shifters, disposed at regular intervals along
the channel, in analogy to the schematics of
Fig.\ref{fig:BB_DecouplingGroup}. Given the small thickness of the
optical elements, the time employed by the light to pass through
them is extremely small and well approximates the BB condition of
instantaneous interaction.

\subsection{Parity-kicks decoupling}
In Ref.~\cite{Vitali1999} it is showed how to inhibit decoherence
through the application of shaped time-varying controls designed
to be equivalent to the application of the parity operator on the
system. The decoupling is realized by a sequence of very frequent
``parity kicks'' (PK) of duration $\tau_0$ after a free evolution
of duration $\Delta t$, $U(\Delta t) = \mathbb{U}_{\Delta t}$.
Assuming that the external pulse is so strong that it is possible
to neglect the free evolution during the pulse $\tau_0$, and that
the pulse Hamiltonian $H_k$ and the pulse width $\tau_0$ can be
chosen to satisfy the PK condition
\begin{equation}
    U_k(\tau_0) \simeq {\rm e}^{-\frac{i}{\hbar}H_k\tau_0} = \mathbb {P}\,,
\end{equation}
the stroboscopic dynamics of the system inhibits decoherence
whenever the following general conditions hold:
\begin{equation}
    \mathbb {P} H_S \mathbb {P} = H_S\hspace{1.5cm}
    \mathbb {P} H_{SE} \mathbb {P} = - H_{SE}.
\end{equation}
This means that the system Hamiltonian is parity invariant and the
interaction with the external environment anticommutes with the
system parity operator.
Decoupling is then obtained through symmetrization with respect to
the group
\begin{equation}
    \mathcal{G} = \{\mathbb {I}, \mathbb {P}\}\,,
\end{equation}
so that $|\mathcal{G}| = 2$.
A harmonic oscillator with linear dissipation
\begin{equation}
    H_{SE} = a^\dag\Gamma+ a\Gamma^\dag\,,
\end{equation}
satisfies such conditions, and the BB control is obtained by
shifting the oscillation frequency $\omega_0$ of $\delta \omega$
during the time $\tau_0$, under the constraint $\delta \omega\,
\tau_0 = \pi$, for which $\mathbb {P} = \exp\{i\pi a^\dag a\}$:
\begin{equation}
    U_k^\dag(\tau_0)\mathbb{U}_{\Delta t}U_k(\tau_0)\,\cdot\,
            \mathbb{I}\,\mathbb{U}_{\Delta t}\, \mathbb{I}\simeq
            \mathbb{P}\, \mathbb{U}_{\Delta t}\, \mathbb{P}\, \mathbb{U}_{\Delta t}\,.
\end{equation}
PK control has a strong relationship with the spin-echo technique,
because in both cases the applied pulses realize a time reversal
of the unwanted Hamiltonian interaction and cancel its effect.

\subsection{Carr--Purcell decoupling for $\mathcal{H}^2$ system}
We consider now a particular system-environment interaction in
which the decoherence occurs along a preferred axis, e.g., along
$z$. In the standard spin-boson model this is described by the
following interaction Hamiltonian:
\begin{equation}\label{eq:Carr-Purcell_Interaction}
    H_{SE} = \sigma_z B_z\,.
\end{equation}
Such an interaction can describe the above-mentioned pure
dephasing process related to $T_2$ in NMR~\cite{Viola2004}, when
the magnetic field is aligned in the $z$ direction, or the
birefringence experienced by a light pulse traveling in a
polarization-maintaining (PM) optical fiber with the fast axis
parallel to the $z$ direction. The ``fast'' and ``slow'' axes are
the PM fiber eigenmodes. The pulses linearly polarized along the
fast axis have a group velocity higher than those polarized along
the slow axis. This implies that a pulse which is not polarized
along one of these directions is split in two separate time bins
with orthogonal polarizations. Such an effect must not be confused
with the polarization-mode dispersion (PMD) of a single-mode (SM)
optical fiber, which is mainly related to a non-perfect symmetry
of the fiber core.

It is possible to use BB controls to reduce the decoherence
induced by the interaction Hamiltonian of
Eq.\ref{eq:Carr-Purcell_Interaction}. Following
Ref.~\cite{Massar2007} a dynamical BB decoupling is realized by
the repetition of only two steps: i) the free evolution of the
system for an interval $\Delta t$ represented by the unitary
$\mathbb{U}_{\Delta t}$; ii) the flip of the qubit around the
$x$-axis through the operator $\mathbb{X}$. After two repetitions
of these basic steps the system dynamics is driven by the
following sequence of operators:
\begin{equation}
    \mathbb{X}\, \mathbb{U}_{\Delta t}\, \mathbb{X}\, \mathbb{U}_{\Delta t} =
        \mathbb{U}_{\Delta t}^\dag \mathbb{U}_{\Delta t} =
        \mathbb{I}\,,
\end{equation}
where the first equality comes from the relation
$\mathbb{X}\mathbb{Z} \mathbb{X} = -\mathbb{Z}$ and where
$\mathbb{X} = \sigma_x$, $\mathbb{Z} = \sigma_z$. The above
sequence $\mathbb{U}_{\Delta t} / \mathbb{X}$ defines a particular
case of dynamical decoupling called ``Carr--Purcell decoupling''
(CP). When the interaction term is like in
Eq.\eqref{eq:Carr-Purcell_Interaction}, i.e., when there is a
preferred basis for decoherence to occur (the ``pointer
basis''~\cite{Zurek2003}) then the CP alone is sufficient to
decouple the system from the environment. More generally, the CP
works perfectly when the noise acts along any axis orthogonal to
the direction of the flip.
Intuitively the spin-flip put forward by the $\mathbb {X}$
operation changes the sign of the interaction Hamiltonian,
Eq.\eqref{eq:Carr-Purcell_Interaction}, which is then symmetrized
with respect to the group
\begin{equation}
    \mathcal{G} = \{\mathbb {I},\mathbb{X}\}\,.
\end{equation}
This averages out the detrimental influence of the environment.

It is worthwhile to mention that the CP decoupling can be easily
applied to PM fibers. It suffices to connect two PM fibers of
equal length in a cross configuration, so that the fast axis of
the first fiber is aligned to the slow axis of the second. The
cross-connector executes an operation analogous to the
$\mathbb{X}$-flip just described, thus removing the difference in
the group velocities due to the fiber birefringence.

\subsection{General decoupling for $\mathcal{H}^2$ system}
In the spin-boson model the most general interaction of a qubit
with its environment is given by
\begin{equation}\label{eq:general_Interaction}
    H_{SE} = \sum_{\alpha =1}^{3} \sigma_\alpha B_\alpha\,.
\end{equation}
Decoherence acts along the axis parallel to the effective magnetic
field $B_{\alpha}$, which is however generally unknown. Therefore
the simple CP-sequence is no more effective. Nonetheless
decoupling can still be obtained through symmetrization with
respect to the complete \textit{Pauli group} PG:
\begin{equation}
    \mathcal{G} = \{{\mathbb {I},\mathbb {X}, \mathbb {Y}, \mathbb {Z}}\}\,,
\end{equation}
with $(\mathbb {X}, \mathbb {Y}, \mathbb {Z}) = (\sigma_x,
\sigma_y, \sigma_z)$ and $|\mathcal{G}| =
4$~\cite{Duan1999,Viola99}. This can be implemented using only two
spin-flips around two orthogonal axes, $\mathbb {X}$ and $\mathbb
{Z}$, because, apart an overall phase, we can write:
\begin{eqnarray}
    \mathbb{Z}\, \mathbb{U}_{\Delta t}\, \mathbb{Z}\,\cdot\,
    \mathbb{Y}\, \mathbb{U}_{\Delta t}\, \mathbb{Y} \!\!\!\!\!&  \cdot   \!\!\!\!\!&
    \mathbb{X}\, \mathbb{U}_{\Delta t}\, \mathbb{X}\,\cdot \,
    \mathbb{I}\,\mathbb{U}_{\Delta t}\,\mathbb{I} \nonumber\\
        &=&
         \mathbb{Z}\, \mathbb{U}_{\Delta t}\,
         \mathbb{X}\, \mathbb{U}_{\Delta t}\,
         \mathbb{Z}\, \mathbb{U}_{\Delta t}\,
         \mathbb{X}\, \mathbb{U}_{\Delta t}\,.
\end{eqnarray}
With PG is possible to decouple the system from the environment,
whatever the direction of the interaction Hamiltonian, up to the
first order in $\omega_CT_C$~\cite{Duan1999,Viola99,Massar2007}.
However it is worth to notice that more general decoupling schemes
are possible which achieve decoupling at higher orders in
$\omega_CT_C$~\cite{Viola2005}.
\begin{figure}[!ht]
   \centering
   \includegraphics[width=.4\textwidth]{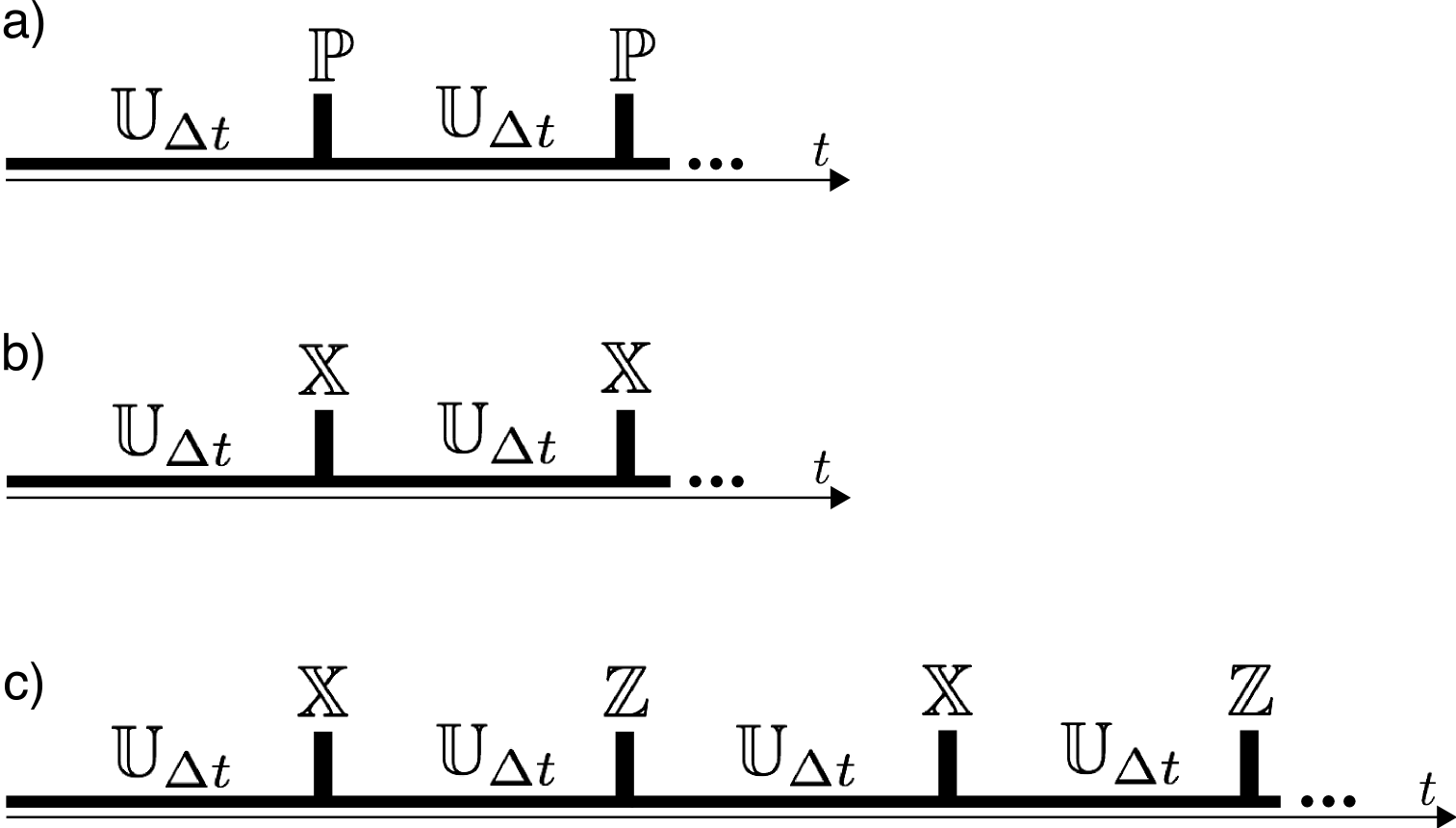}
   \caption{Decoupling BB cycles for: a) parity-kicks (PK); b) Carr--Purcell
   (CP); c) Pauli group.}
   \label{fig:BB_DecouplingExamples}
\end{figure}

Fig.\ref{fig:BB_DecouplingExamples} summarizes the open-loop
decoupling techniques described so far, in the BB approximation of
instantaneous control pulses. The inset (a) and (b) of the figure
contain the schematics of PK and CP decoupling techniques
respectively; their use in the optical domain has already been
discussed. The inset (c) represents the most general decoupling
scheme, the PG. It becomes particularly important when the
preferred axis of the system-environment interaction is unknown,
e.g., in an SM optical fiber. In fact, it is possible to segment
an SM fiber into smaller and smaller portions until a unity of
length $L_0$ is reached which features a constant birefringence
(along unknown axes) and so constant dispersive properties.
Regardless of the direction of the birefringence, PG decoupling
can be applied on a length scale comparable to $L_0$ to suppress
decoherence on each segment of the SM fiber and, by consequence,
on the fiber as a whole.

\subsection{Decoherence for polarization-qubits}
All of the described decoupling scheme, in particular the PG, are
suitable for a straightforward application in the optical domain.
Photons are ideal carriers of quantum information as they allow a
fast transfer of information while interacting weakly with the
surroundings. However in the practice photons are replaced by
light pulses which are intrinsically non-monochromatic and are
measured by detectors which do not possess an infinite integration
time. Hence, a vast part of the information committed to the
frequency degree of freedom is traced out by the final measuring
process, which is not able to discriminate each frequency
component of the light pulse.

A solution to this problem is to encode information in the
polarization degree of freedom. However, when the light propagates
in a dispersive medium, such as an optical fiber, the initial
information can spread out to other unwanted degrees of freedom,
like frequency, and again that information will go lost. So in
this case the frequency degree of freedom acts as the environment
of the polarization qubit and induces a dephasing process on it.

This happens because the dispersive medium couples the two degrees
of freedom with an interaction Hamiltonian like the following:
\begin{equation}
    H_{SE} =  \sum_{\alpha =1}^{3} \sigma_\alpha n_\alpha(\omega)\,,
    \label{eq:birefrHamiltonian}
\end{equation}
where $n_\alpha(\omega)$ is the frequency-dependent refractive
index of the medium and $\sigma_\alpha$ are the generators of the
polarization-qubit evolution. The above Hamiltonian descends from
the Fourier transform of the one describing a light pulse
propagating in an inhomogeneous medium, i.e. with an index of
refraction that depends on the position of the pulse. So it is
closely related to the above-mentioned PMD occurring in SM optical
fibers. The similarity with Eq.\eqref{eq:general_Interaction} is
apparent; this means that PG can be employed to adverse the
polarization decoherence due to PMD. In a classical communication
scenario PMD adversely affect the rate of long distance
communications~\cite{Poole88,Galtarossa2000}. PMD is also
substantial in a fiber-based quantum communication if one uses a
qubit realized in polarization since the unavoidable coupling with
the birefringent environment affect the purity and fidelity of
transmitted qubits~\cite{Gisin2002}

There are not many techniques to combat the PMD detrimental
effects. One is of course to adopt a PM fiber, but it works only
for the two linear polarizations aligned to the fiber eigenmodes.
Another technological solution is to spin the SM fiber so to
distribute inhomogeneities symmetrically in the
core~\cite{Barlow1981};  however this increases the fiber
attenuation more than 20 times for wavelengths in the third
Telecom window. Finally, one solution is based on Faraday
mirrors~\cite{Martinelli1989, Berglund2000}. The principle is that
of ``retracing beam'': a light pulse travels back and forth along
the communication channel; this leads it to compensate during the
backward travel all the PMD encountered in the forward travel. The
only assumption in this case is that the PMD remains constant
during the two trips of the pulse~\cite{Martinelli1989}. This
technique is clearly not applicable to all those channels where
the information carriers travel in the forward direction only.

Given the lack of conclusive solutions to adverse the PMD on
one-way communication channels, it is quite surprising that BB
dynamical decoupling has not been already applied to the optical
domain. In fact specific experiments on this subject are missing.
The remarkable experiments reported in~\cite{Berglund2000} cannot
be deemed an exception in this respect, since they {\bl
partly} 
rely on the principle of retracing beam; so their application {\bl
appears to be more suitable for} two-way communication channels.
In the following we focus on the recent demonstration given
in~\cite{Damodarakurup2009} and show the effectiveness of the BB
dynamical decoupling for suppressing the decoherence of a
polarization qubit confined in a \textit{ring cavity}. The choice
of a ring cavity has a twofold motivation: on one side it realizes
the Hamiltonian of Eq.\eqref{eq:birefrHamiltonian}, which well
mimic the effect of PMD in a dispersive medium. The second reason
is that a ring cavity represents the perfect simulation of a
one-way channel: the principle of retracing beam is removed as its
root because errors tend to accumulate rather than to cancel out.

\section{The experimental setup and its characterization}
\begin{figure}[!ht]
   \centering
   \includegraphics[width=.475\textwidth]{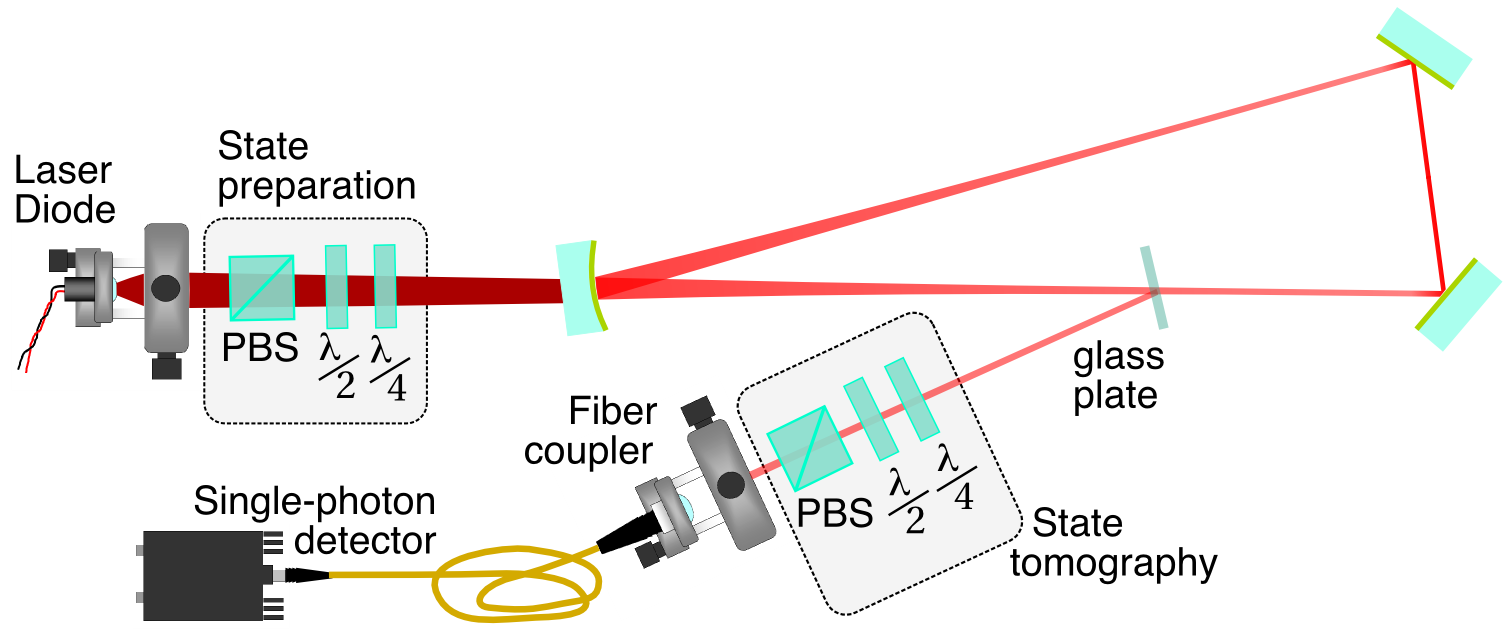}
   \caption{(Color online) Schematics of the experimental apparatus. The polarization
   of pulsed laser diode is determined by a state preparation apparatus, which is constituted
   by a polarizing beam-splitter (PBS), a half-wave plate ($\lambda/2$) and a quarter-waveplate
   ($\lambda/4$). The laser pulses are attenuated and injected in a triangular-ring cavity
   constituted by a spherical mirror and two plane mirrors at 45$^\circ$. The pulses trapped
   in the cavity are extracted by means of a glass plate and sent into the state tomography
   apparatus constituted by a PBS, $\lambda/2$ and $\lambda/4$. The light is finally coupled
   to an optical fiber and detected by a single-photon avalanche photodiode.}
   \label{fig:BB_Setup_OnlyCavity}
\end{figure}

The experimental apparatus of Ref.\cite{Damodarakurup2009} is
depicted in Fig.~\ref{fig:BB_Setup_OnlyCavity}. A laser diode with
central wavelength at $\lambda_0 \simeq 800$~nm and bandwidth
$\Delta \lambda \simeq 15$~nm is pulsed at repetition rate of
100~KHz and pulse duration $\sim$100~ps. The laser is attenuated
and injected in a triangular ring cavity through a spherical
mirror.
The polarization state of the laser is prepared by using a
polarizing beam-splitter (PBS), a half-wave plate ($\lambda/2$)
and a quarter-wave plate ($\lambda/4$) with a very high degree of
purity.
At every round trip the light is extracted from the cavity with
$4\%$ probability using a $\sim$100~$\mu$m thin glass plate. Its
polarization is analyzed by means of the tomographic
technique~\cite{James2001}, and then sent into a multimode fiber
connected to a single-photon detector with quantum efficiency
$\sim 70\%$.
Let us now separately discuss the generation, processing, and
measurement of the qubit.

\subsection{Operational polarization qubit}
The attenuated pulse entering the cavity through the input
spherical mirror and after the state preparation stage can be well
described as a multi-mode weak coherent state, which can be
written as
\begin{eqnarray}\label{eq:input-weak-cohe-state}
     |\psi\rangle_{in} = \exp\left\{\int d\omega \left[\alpha_H(\omega)
     \hat{a}_H(\omega)^{\dagger}+\alpha_V(\omega) \hat{a}_V(\omega)^
     {\dagger}\right.\right. \nonumber\\
     \left.\left.\pm H.C.\right]\right\}|vac\rangle\,.\,\,
\end{eqnarray}
$\hat{a}_S(\omega)^{\dagger}$ is the operator creating a photon
with polarization $S=\{H,V\}$ and frequency $\omega$, satisfying
the commutation relation
$\left[\hat{a}_S(\omega),\hat{a}_{S'}(\omega')^{\dagger}\right]=\delta_{S,S'}\delta(\omega-\omega')$,
and $|vac\rangle $ denotes the vacuum state. We denote with $V$
($H$) the polarization orthogonal (parallel) to the plane of the
cavity.

The polarization of the pulse is fully characterized by its Stokes
parameters, which are determined by the functions
$\alpha_H(\omega)$ and $\alpha_V(\omega)$ according to
\begin{eqnarray}\label{eq:stokes0}
    s_0 \!\!&=&\!\!\! \int d\omega \left[|\alpha_H(\omega)|^2+ |\alpha_V(\omega)|^2\right], \\
    s_1 \!\!&=&\!\!\! \int d\omega \left[|\alpha_H(\omega)|^2- |\alpha_V(\omega)|^2\right], \\
    s_2 \!\!&=&\!\!\! \int d\omega \left[\alpha_H(\omega)^*\alpha_V(\omega)+\alpha_H(\omega)
    \alpha_V(\omega)^*\right],\\
    s_3 \!\!&=&\!\!\! -i\int d\omega \left[\alpha_H(\omega)^*\alpha_V(\omega)-\alpha_H(\omega)
    \alpha_V(\omega)^*\right]. \label{eq:stokes3}
\end{eqnarray}
One can associate to the full state of the field of
Eq.~(\ref{eq:input-weak-cohe-state}) a polarization state
described by the $2\times 2$ qubit density matrix
\begin{equation}\label{eq:rhopolarization}
    \rho_{pol}=\frac{1}{2}(1+\vec{P}\cdot \vec{\sigma}),
\end{equation}
where $\vec{\sigma}=(\mathbb{X},\mathbb{Y},\mathbb{Z})^T$ is the
vector of the three Pauli matrices, and
$\vec{P}=(s_2/s_0,s_3/s_0,s_1/s_0)^T$ is the Bloch vector,
completely characterizing the state of the polarization qubit. The
Stokes parameters of Eqs.~(\ref{eq:stokes0})-(\ref{eq:stokes3})
satisfy the condition $s_0^2-s_1^2-s_2^2-s_3^2 \geq 0$, which is
equivalent to the condition on the Bloch vector norm
$|\vec{P}|\leq 1$, which is required by the fact that $\rho_{pol}$
must describe a physical density matrix. This condition can be
verified using the fact that
Eqs.~(\ref{eq:stokes0})-(\ref{eq:stokes3}) imply
\begin{eqnarray}\label{eq:stokes-cond}
    &&s_0^2-s_1^2-s_2^2-s_3^2 \\
    &&=\int \int d\omega d \omega'\left|\alpha_H(\omega)\alpha_V(\omega')-\alpha_H(\omega')\alpha_V(\omega)\right|^2 \geq
    0. \nonumber
\end{eqnarray}
Notice that this latter equation shows that the inequality becomes
an equality, i.e., $|\vec{P}|= 1$ and the polarization state is a
pure qubit state, if and only if
$\alpha_H(\omega)\alpha_V(\omega')=\alpha_H(\omega')\alpha_V(\omega)$,
$\forall \omega,\omega'$, which is realized if and only if the
ratio $\alpha_H(\omega)/\alpha_V(\omega)$ does not depend upon
$\omega$. This is verified when the polarization state of the
pulse is independent of its frequency distribution, i.e.,
polarization and frequency are factorized and one can write
$\alpha_S(\omega)=\epsilon(\omega) \alpha_S$ ($S=H,V$). Therefore
for the multi-mode coherent state of
Eq.~(\ref{eq:input-weak-cohe-state}), the polarization state
$\rho_{pol}$ is pure if and only if is ``disentangled'' from the
frequency $\alpha_S(\omega)=\mathcal{E}(\omega) \alpha_S$
($S=H,V$), with $\mathcal{E}(\omega)$ the amplitude spectrum of
the pulse normalized such that $\int d\mu_\omega\equiv\int d\omega
|\mathcal{E}(\omega)|^2 = 1$, and in this case it is represented
by a frequency-independent Jones vector
\begin{equation}
    |\pi\rangle_{in} =\frac{1}{\sqrt{|\alpha_H|^2+|\alpha_V|^2}}
        \left(
        \begin{array}{c}
            \alpha_H\\
            \alpha_V
        \end{array}
            \right)\,.
\end{equation}
On the contrary, if $\alpha_H(\omega)/\alpha_V(\omega)$ depends
upon $\omega$, the Jones vector is frequency-dependent, and
$\rho_{pol}$ is mixed: polarization and frequency are effectively
``entangled''.

The weak coherent pulse is then injected into the cavity, where it
is transformed by the action of mirrors, which are linear and
passive elements and therefore can only mix
$\hat{a}_H(\omega)^{\dagger}$ and $\hat{a}_V(\omega)^{\dagger}$,
without involving processes like creation or annihilations of
photons. Also the additional elements which will be later added
within the cavity for the realization of the BB control, i.e.,
wave-plates and Soleil--Babinet(S-B) compensators, are linear and
passive devices, and therefore the action of a generic element
within the cavity can always be described in terms of a $2\times 2
$ unitary matrix form. The overall transformation for each
round-trip in the cavity is given by:
\begin{equation}\label{}
    \left[
    \begin{array}{c}
    \hat{a}_H(\omega)^{\dagger}\\
    \hat{a}_V(\omega)^{\dagger}
    \end{array}
    \right]
    \rightarrow \mathbf{U}(\omega)
    \left[
    \begin{array}{c}
    \hat{a}_H(\omega)^{\dagger}\\
    \hat{a}_V(\omega)^{\dagger}
    \end{array}%
    \right]\,,
\end{equation}
and the coherent state pulse is changed into
\begin{eqnarray}\label{expa}
    |\psi\rangle_{tr} = \exp\left\{\int d\omega \left[\beta_H(\omega)\hat{a}_H(\omega)^{\dagger}
            +\beta_V(\omega)\hat{a}_V(\omega)^{\dagger}\right.\right. \nonumber\\
         \left.\left.\pm H.C.\right]\right\}|vac\rangle\,,\,
\end{eqnarray}
where
\begin{equation}\label{}
    \left(
    \begin{array}{c}
        \beta_H(\omega)\\
        \beta_V(\omega)^{\dagger}
    \end{array}%
    \right) =\mathbb{U}(\omega)^T
     \left(
    \begin{array}{c}
        \alpha_H(\omega)\\
        \alpha_V(\omega)
    \end{array}%
    \right).
\end{equation}
The Stokes parameters and consequently the polarization state is
changed accordingly.

The final attenuator (in the experimental setup the glass plate),
$\eta$, decreases the intensity of the pulse, $\beta_S(\omega) \to
\eta \beta_S(\omega) $ with $\eta \ll 1$. Assuming $\eta^2\int
d\omega|\beta_S(\omega)|^2\ll 1$, the state of the pulse at the
cavity output and just at the entrance of the detection stage can
be well approximated by the first order linear expansion of the
exponential in Eq.~(\ref{expa})
\begin{eqnarray}\label{lin}
    |\psi\rangle_{tr} \simeq |vac\rangle &\\
    +  \eta \int d\omega \!\!&\!\!\left[\beta_H(\omega)\hat{a}_H(\omega)^{\dagger}
     +\beta_V(\omega)\hat{a}_V(\omega)^{\dagger}\right]|vac\rangle\,. \nonumber
\end{eqnarray}
By post-selecting the events with at least one photon we discard
the vacuum component and the state entering the detector is well
approximated by a single-photon state given by the integral term
on the right hand side of Eq.~(\ref{lin}).

One issue that should be addressed is the use of the term
``polarization qubit'' for the description of the laser pulse
traveling in the ring cavity just described.
If the frequency distribution is independent of the polarization,
one can write $\beta_S(\omega)=\, \mathcal{E}(\omega) \beta_S$
($S=H,V$), and factorize the frequency and polarization degree of
freedom
\begin{equation}\label{single}
    |\psi\rangle_{out} \simeq  \int d\omega \, \mathcal{E}(\omega)|\omega \rangle \left[\beta_H |H\rangle +\beta_V|V\rangle\right].
\end{equation}
If instead the polarization state is not independent of the
frequency, the two degrees of freedom are entangled and one has
decoherence when one looks at the polarization state only.
What is relevant is that the Bloch vector $\vec{P}$ and therefore
the polarization state is not changed by the extraction by the
glass plate, and, obviously, by the post-selection as well.
Therefore the output single photon state, ``operational''
polarization qubit, perfectly describes the polarization state
within the cavity, even if we have a coherent state pulse within
the optical system with a non-zero probability of having more than
one photon.

In our experiment the state emerging from the laser is a coherent
state with average photon number per pulse about equal to 1 soon
after the mirror. The resulting coherent state entering the
detection apparatus has a mean photon number per pulse  $\mu \leq
4\times10^{-2}$, and the probability of having two or more photons
in a detection event is less than 2\%, as we have experimentally
verified.
The experimental setup detailed above then justifies a description
in terms of single-photon polarization qubits, and allows us to
write the state injected into the cavity, soon after the spherical
mirror, as an effective single-photon state
\begin{equation}\label{single}
    |\psi\rangle^{\rm eff}_{in} =
        \int d\omega\, \mathcal{E}(\omega)|\omega \rangle\otimes|\pi\rangle_{in}\,,
\end{equation}
where $|\omega \rangle\otimes|\pi_{in}\rangle =
[\alpha_H\hat{a}_H(\omega)^{\dagger}+\alpha_V\hat{a}_V(\omega)^{\dagger}]|0\rangle$.

\subsection{The cavity}
\begin{figure}[!ht]
   \centering
   \includegraphics[width=.375\textwidth]{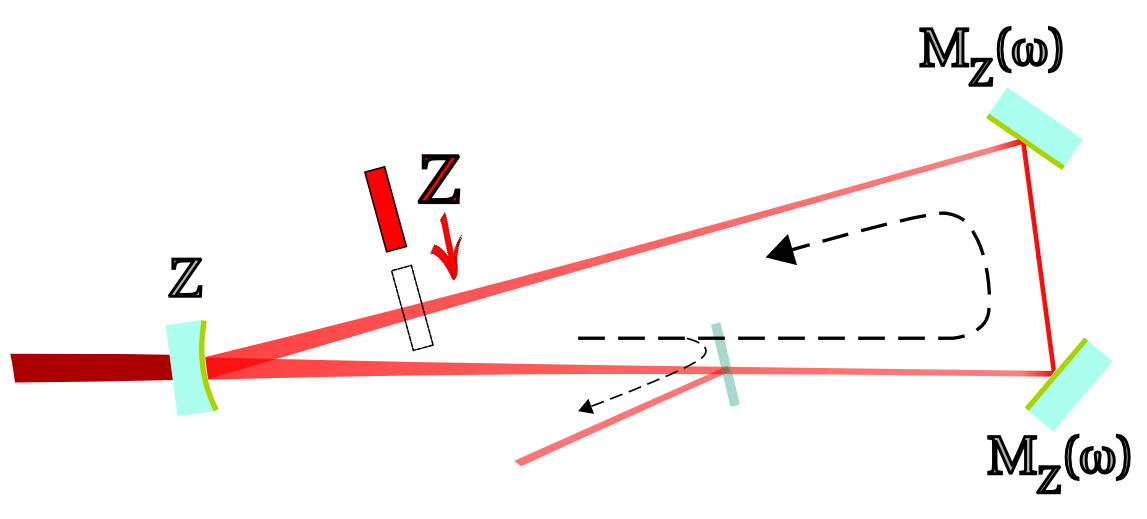}
   \caption{(Color online) Schematics of the triangular-ring cavity
   constituted by a spherical mirror and two plane mirrors at 45$^\circ$,
   which realize on the polarization a unitary operation, $\mathbb{Z}$,
   and a frequency dependent unitary operations, $\mathbb{M_Z}(\omega)$,
   respectively. A wave-plate inserted in a long arm of the cavity implements
   the $\mathbb{Z}$ operation in order to compensate for the spherical mirror
   $\mathbb{Z}$ transformation.}
   \label{fig:BB_Cavity_Z}
\end{figure}

The triangular ring cavity (see Fig.~\ref{fig:BB_Cavity_Z}) is
realized by a spherical mirror with radius of curvature 1~m and
reflectivity $\sim$~98\%. The cavity is also formed by two flat
mirrors at $45^{\circ}$ with reflectivity $\gtrsim$~99\%.  The
aperture angle of the cavity at the spherical mirror is
$\sim8^\circ$. The long arms of the cavity are $\sim 0.94$~m and
the short arm is $\sim 0.13$~m determining a total cavity length
of $\sim2.01$~m.
A cavity mirror generally produces a phase shift which depends
upon both frequency and polarization, and therefore its action is
described by the matrix~\cite{Azzam1977} (keeping always the
right-hand coordinate system before and after
reflection~\cite{Brinkmeyer1981,SalehTeich})
\begin{equation}\label{}
  \mathbb{M_Z}(\omega)  =
   \left[
    \begin{array}{cc}
    \rm{e}^{-i\phi_H(\omega)} & 0\\
    0 & e^{-i\phi_V(\omega)+i\pi}
    \end{array}
   \right]
    = \mathbb{Z}\exp\{-i\phi(\omega)\mathbb{Z}\},
\end{equation}
where we have omitted an unessential global phase factor and we
have defined with $\phi(\omega)= \phi_H(\omega)-\phi_V(\omega)$
the relative phase due to the polarization-reflectivity
difference. This is due to the different reflection coefficients
for orthogonal and parallel polarizations with respect to the
plane of the cavity (s- and p- polarization, respectively). The
two plane mirrors at 45$^\circ$ can be assumed identical and are
therefore characterized by the same matrix $\mathbb{M_Z}(\omega)$.
The third concave mirror of the cavity is almost at normal
incidence: this mirror does not distinguish $H$ and $V$
polarization, so that $\phi(\omega)\simeq 0$ and therefore it acts
as $\mathbb{M_Z}(\omega)\simeq \mathbb{Z}$.
The polarization qubit in the cavity is subjected to a phase-noise
along $z$-axis and the unitary operator describing the
polarization transformation after one cavity round-trip for a
given frequency component is given by
\begin{eqnarray}\label{eq:OnlyCavityTransformation}
    \mathbb{U}[\phi(\omega)]
        &=& \mathbb{Z}\cdot\mathbb{M_Z}(\omega)\cdot\mathbb{M_Z}(\omega)\nonumber\\
        &=& \mathbb{Z}\exp\{-i2\phi(\omega)\mathbb{Z}\}\,,
\end{eqnarray}
The dispersive properties of the two plane mirrors at
$45^{\circ}$, i.e., their frequency dependence, generates
polarization decoherence: in fact the frequency dependence of the
matrix $\mathbb{M_Z}$ appears only if the optical element is
dispersive,
otherwise it is constant and it does not entangle polarization and
frequency degrees of freedom.

\subsection{State tomography reconstruction}
At every round trip the light is extracted from the cavity using
the thin glass plate as shown in the setup of
Figs.~\ref{fig:BB_Setup_OnlyCavity} and~\ref{fig:BB_Cavity_Z}. The
output polarization state after $n$ round-trips is given by the
reduced density matrix obtained by tracing over the frequency
degree of freedom
\begin{equation}\label{eq:rho_n}
    \hat{\rho}_{out}^{(n)} = \int d\mu_\omega
        \mathbb{U}[\phi(\omega)]^n |\pi\rangle_{in}\langle\pi| \mathbb{U}^\dag[\phi(\omega)]^n\,.
\end{equation}

\begin{figure}[!ht]
   \centering
   \includegraphics[width=.475\textwidth]{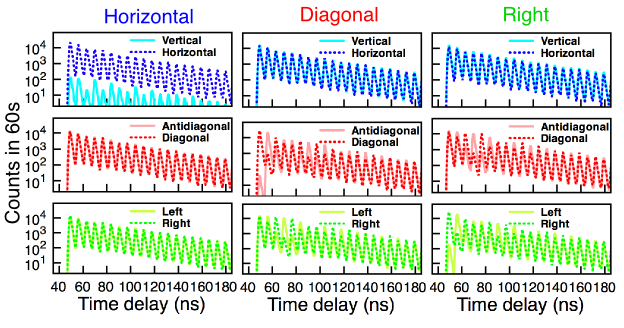}
   \caption{(Color online) Typical acquisition runs, with a sequence
   of peaks each corresponding to a cavity round-trip, is shown for a
   \textit{horizontal}, \textit{diagonal}, and \textit{right} polarization
   input states. The interval between adjacent peaks amounts to 6.80~ns,
   in agreement with the given cavity-length. The three plots correspond
   to the measurements in the bases, from the top, $\mathbb{Z}$, $\mathbb{X}$
   and $\mathbb{Y}$, respectively. It is worth to note that for the horizontal
   polarization input state the vertical contribution is negligible and the
   outputs in the $\mathbb{X}$ and $\mathbb{Y}$ are balanced, as expected for
   an eigenstate of the cavity unitary evolution.}
   \label{fig:PLOT_HDR_Log}
\end{figure}

Experimentally the output density matrix after $n$ round-trips has
been evaluated by the acquisition of the  output signal from the
detector, which stops the time conversion in a Time-to-Amplitude
Converter (TAC) synchronized with the laser. The time delay
between the input and output signals is then recorded by a
Multi-Channel Analyzer (MCA) with a resolution of 8192 channels.
The acquisition electronics has a time-resolution of $102$~ps.
Typical acquisition runs are shown in Fig.~\ref{fig:PLOT_HDR_Log},
showing a sequence of peaks, each corresponding to a cavity
round-trip. The interval between adjacent peaks amounts to
$\sim$6.80~ns, in agreement with the measured length of the
cavity. In all the experiments both the decoherence and the effect
of BB control are studied by sending specific input polarization
in the cavity and capturing the photon round trip from the cavity
with respect to different tomographic measurement bases.
Fig.~\ref{fig:PLOT_HDR_Log} depicts the photon round trips for
\textit{horizontal} (H), \textit{diagonal} (D), and \textit{right}
(R) polarization with respect to $\mathbb{Z}$, $\mathbb{X}$ and
$\mathbb{Y}$ basis tomographic measurement. The decay of peaks
after every round trip for H polarization input state is highly
uniform with respect to all basis measurements but for D and R
polarizations it is highly non-uniform, indicate that the
eigenstates of the cavity unitary evolution are H and V
\textit{vertical} polarizations.

\begin{figure}[!ht]
   \centering
   \includegraphics[width=.475\textwidth]{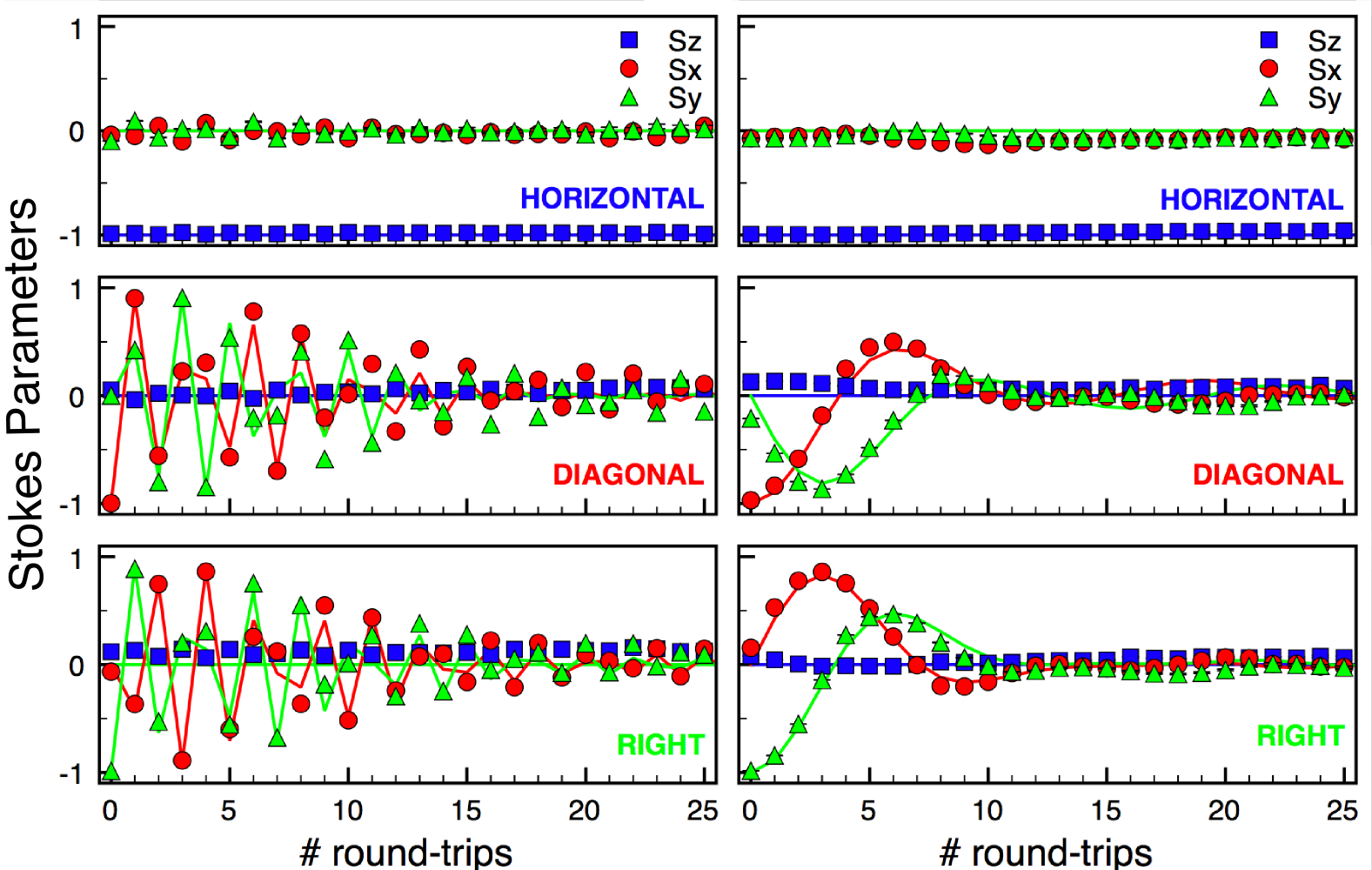}
   \caption{(Color online) Stokes' parameters as a function of the
   number of round-trips of the cavity, for  \textit{horizontal},
   \textit{diagonal}, and \textit{right} polarization input states:
   left) only cavity; right) cavity with a $\mathbb{Z}$ wave-plate
   in a long arm in order to compensate for the spherical mirror
   $\mathbb{Z}$ transformation, as shown in Fig.~\ref{fig:BB_Cavity_Z}.
   Note how the wave-plate inserted in a long arm of the cavity cancels
   the polarization-flip for diagonal and right polarization input states
   due to the spherical mirror transformation. The wave-plate has no effect
   instead in the case of a horizontal polarization input state, which is
   an eigenstate of the cavity transformation in both cases.
   The lines following the experimental data are evaluated by means of a
   numerical simulation with two fitting parameters: the standard deviation
   of the Gaussian measure $\sigma_\phi^{\rm est}  = 8.39\times10^{-2}\,{\rm rad}$,
   and the phase difference for horizontal and vertical polarization, $\phi_0 = -0.2182\,{\rm rad}$.}
   \label{fig:BB_SVector_Compare_OnlyCav_longZ}
\end{figure}

This is made evident by the time evolution of the Stokes
parameters for D and R input polarization rates, as shown by the
left panels in Fig.~\ref{fig:BB_SVector_Compare_OnlyCav_longZ}.
The Stokes' parameters are calculated as expectation values of the
Pauli matrices ($\sigma_z, \sigma_x, \sigma_y$) over the density
matrices evaluated by maximum-likehood optimization
procedure~\cite{James2001} on the six experimental counts obtained
by integration of ten time bins around each peak for the three
polarization measurement bases.

A numerical simulation of Eq.~\eqref{eq:rho_n} has been run
assuming a Gaussian frequency spectrum for the input pulse given
by
\begin{equation}\label{eq:AmplSpectrum}
    \mathcal{E}(\omega) =
        \frac{1}{\sqrt[4]{\pi\sigma^2_\omega}}\exp[-(\omega-\omega_0)^2/2\sigma_\omega^2],
\end{equation}
where $\sigma_\omega$ represents the bandwidth of the radiation
spectrum centered in $\omega_0$. The frequency dependence of the
integrand is determined by the fact that, with a very good
approximation, the phase shift is a linear function of the
frequency, $\phi \simeq \phi_0 + \tau \omega$. The integral can be
therefore transformed into an average over a Gaussian measure,
$d\mu_\phi$, with standard deviation $\sigma_\phi = \tau
\sigma_\omega$.
The linear passive optical elements constituting the cavity
(mirrors and wave-plates) have been modeled by $2\times2$ unitary
matrices, and the $n$-th round-trip polarization density matrix
has been evaluated by the average over a Monte-Carlo simulation of
the measure $d\mu_\phi$ of the $n$-th power of the matrix
describing the effect of a cavity round-trip.
The lines fitting the experimental data in
Fig.~\ref{fig:BB_SVector_Compare_OnlyCav_longZ} and following, are
the numerical fit with two parameters: the standard deviation of
the Gaussian measure $\sigma_\phi^{\rm est}$, and the phase
difference between horizontal and vertical polarization, $\phi_0$.

\subsection{Polarization decay from the cavity}
In order to understand the polarization decoherence caused by
cavity round-trips, we have first studied the \textit{cavity only}
configuration as shown in Figs.~\ref{fig:BB_Setup_OnlyCavity}
and~\ref{fig:BB_Cavity_Z}. To quantify the decoherence, we use
both purity
\begin{equation}
    \mathcal{P}= Tr(\rho^{2}_{out})\,,
\end{equation}
and fidelity
\begin{equation}\label{eq:fid_def}
    \mathcal{F}=~_{in}\langle \pi|\rho_{out}|\pi \rangle_{in},
\end{equation}
where $\rho_{out}$ is the output density matrix and
$|\pi\rangle_{in}$ is the pure input polarization state. Purity
$\mathcal{P}$ is maximum and equal to one for pure states and to
$1/d$ for maximally mixed states of dimension $d$. Fidelity
$\mathcal{F}$ measures the distance between quantum states and the
definition of Eq.~(\ref{eq:fid_def}) is valid in the case of the
distance between a mixed and a pure state. For two density
matrices $\rho$, $\sigma$ it is generalized to
$F(\rho,\sigma)=(\textrm{tr}\sqrt{\sqrt{\rho}\,\sigma\sqrt{\rho}})^2$~\cite{Jozsa1994}.
However, one can adopt the alternative definition
$F^\prime(\rho,\sigma)=\textrm{tr}\sqrt{\sqrt{\rho}\,\sigma\sqrt{\rho}}$~\cite{NielsenChuang2000},
sometimes denoted as $\sqrt{F}$ and called square root fidelity.
\begin{figure}[!ht]
   \centering
   \includegraphics[width=.475\textwidth]{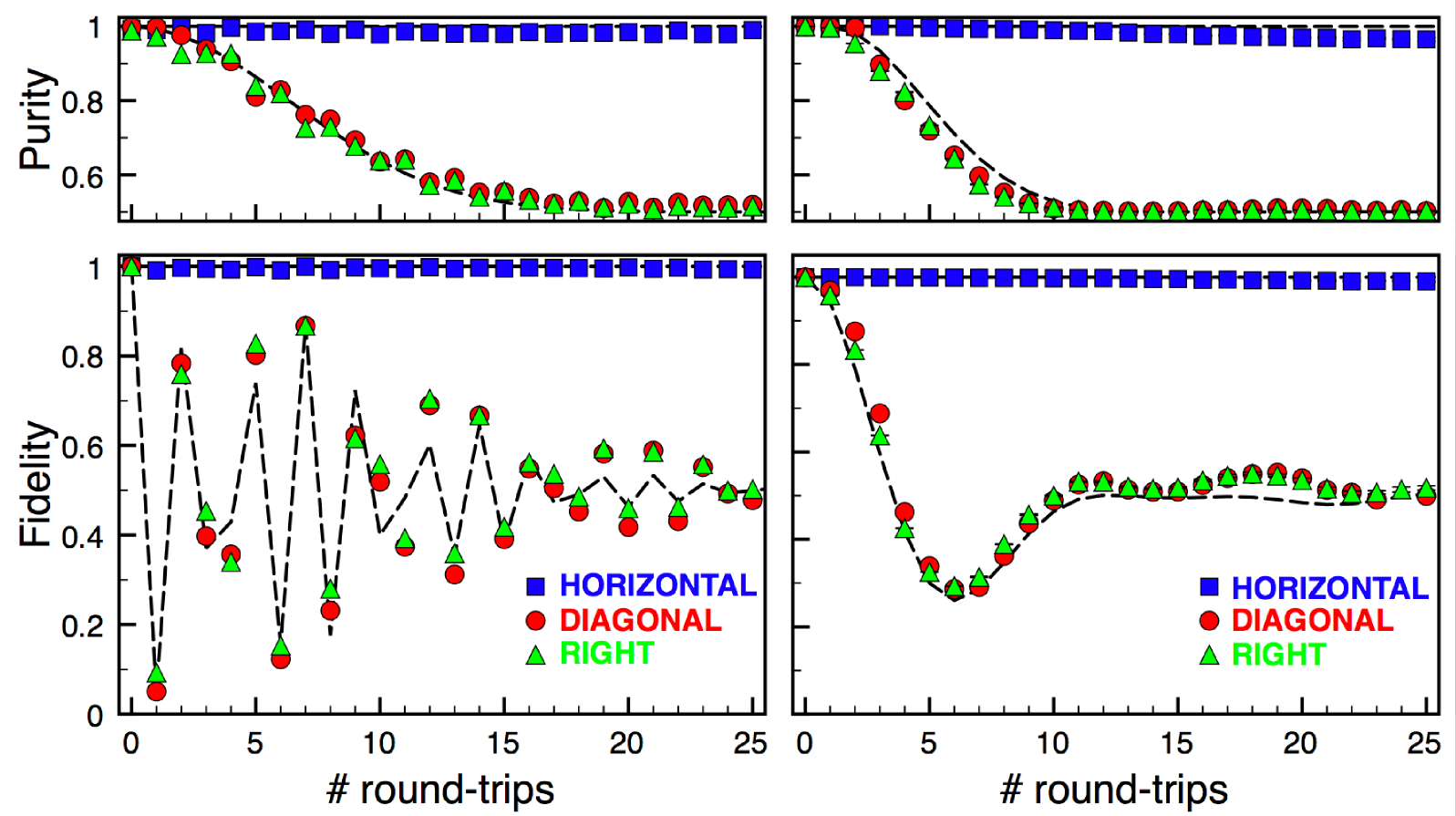}
   \caption{(Color online) Purity and fidelity versus the number of
   round-trips for three different input polarization states: left)
   only cavity; right) cavity with a $\mathbb{Z}$ in a long arm in
   order to compensate for the spherical mirror $\mathbb{Z}$ transformation,
   as shown in Fig.~\ref{fig:BB_Cavity_Z}.  \textit{Horizontal} polarization
   is well preserved. On the contrary the \textit{diagonal} and \textit{right}
   states quickly decay to the fully mixed state (purity and fidelity equal to
   $1/2$). Dashed lines curves are obtained by numerical simulation with the
   same parameters of the previous figure.}
   \label{fig:BB_PurityFidelity_Compare_OnlyCav_longZ}
\end{figure}

The purity and fidelity after each cavity round trip, evaluated
from the experimentally reconstructed density matrices, are shown
on the left panels in
Fig.~\ref{fig:BB_PurityFidelity_Compare_OnlyCav_longZ}. As
expected, the cavity preserves the purity and fidelity only for H
polarization input, while for D and R input states, they both
quickly decay to the fully mixed state ($\mathcal{P}= \mathcal{F}
= 1/2$). Assuming the amplitude spectrum of
Eq.~\eqref{eq:AmplSpectrum}, it is possible to derive analytically
a Gaussian decay of the purity as a function of the number of
round-trips $n$,
\begin{equation}\label{eq:purity}
    \mathcal{P}=\frac{1}{2}[1+\exp(-2n^{2}\sigma^{2}_{\phi})].
\end{equation}
The standard deviation of the Gaussian distribution of the
phase-shift, $\sigma_{\phi}$, is estimated by a best fit on the
experimental data of the purity reported in the left panel of
Fig.~\ref{fig:BB_PurityFidelity_Compare_OnlyCav_longZ} using the
expression of Eq.~\eqref{eq:purity}. The obtained value agrees
with the estimated parameter from the numerical simulation and it
is given by $\sigma_\phi^{\rm est} = 8.39\times10^{-2}\,{\rm
rad}$. Differently from the purity, the fidelity depends upon both
$\sigma_{\phi}$ and the parameter $\phi_0$, which is the phase
difference between H and V polarization reflection at the central
frequency $\omega_0$, as highlighted by the slowly-varying
envelope of the fidelity shown in
Fig.~\ref{fig:BB_PurityFidelity_Compare_OnlyCav_longZ}. The flips
of the fidelity every round-trip for D and R input polarization
state, shown in the left panel of
Fig.~\ref{fig:BB_PurityFidelity_Compare_OnlyCav_longZ}, are
instead due to the spherical mirror transformation $\mathbb{Z}$.
The dashed lines following the purity and fidelity data reported
in Fig.~\ref{fig:BB_PurityFidelity_Compare_OnlyCav_longZ}
represent the numerical fit corresponding to the values
$\sigma_\phi^{\rm est} = 8.39\times10^{-2}\,{\rm rad}$ and $\phi_0
= -0.2182\,{\rm rad}$.

\subsection{Compensating the spherical mirror}
In a single round trip the polarization qubit encounters two
decohering elements (the two plane mirrors represented by
$\mathbb{M}(\omega)$), and a $\mathbb{Z}$ transformation due to
the spherical mirror. As shown by
Eq.~\eqref{eq:OnlyCavityTransformation}, the transformation of the
polarization qubit realized by a cavity round-trip is given by the
product of a phase-shift $\exp\{-i2\phi(\omega)\mathbb{Z}\}$, with
Gaussian-distributed random phase, and a $\mathbb{Z}$ operation,
which flips every round trip the D and R polarization, as
elucidated by the data reported in the left panels of
Figs.~\ref{fig:BB_SVector_Compare_OnlyCav_longZ},
and~\ref{fig:BB_PurityFidelity_Compare_OnlyCav_longZ}.
In order to compensate these $\mathbb{Z}$-flips, we have inserted
in one of the long arms of the cavity, before the spherical
mirror, an additional wave-plate set to realize a further
$\mathbb{Z}$ operation, as shown in Fig.~\ref{fig:BB_Cavity_Z}.
With this new optical element in the setup, the evolution of the
polarization state after a round-trip is given by
\begin{eqnarray}\label{eq:CavityWithZ}
   \mathbb{U}[\phi(\omega)]
       &=& \mathbb{Z}\cdot\mathbb{Z}\cdot\mathbb{M_Z}(\omega)\cdot\mathbb{M_Z}(\omega)\nonumber\\
       &=& \exp\{-i2\phi(\omega)\mathbb{Z}\}\,.
\end{eqnarray}
The behavior of purity and fidelity in this new configuration is
shown in the right panels of
Fig.~\ref{fig:BB_PurityFidelity_Compare_OnlyCav_longZ}. The flips
in the fidelity decay are now completely eliminated and only the
slowly-varying envelop due to $\phi_0$ is left.
As shown in the right panels of
Fig.~\ref{fig:BB_SVector_Compare_OnlyCav_longZ}, also the flips in
the Stokes' parameters for D and R input polarizations are
eliminated in the new cavity configuration.

\section{Bang-bang decoupling}
Here we experimentally demonstrate the application of BB dynamical
decoupling for suppressing decoherence on single-photon
polarization qubits flying in the cavity. Polarization decoherence
takes place in localized spatial regions (mirrors, S--B, ...), and
the BB controls will be implemented in space rather than in time
by placing wave-plates along the photon path.

\subsection{Suppressing polarization decoherence of the ring cavity}
\begin{figure}[!ht]
   \centering
   \includegraphics[width=.375\textwidth]{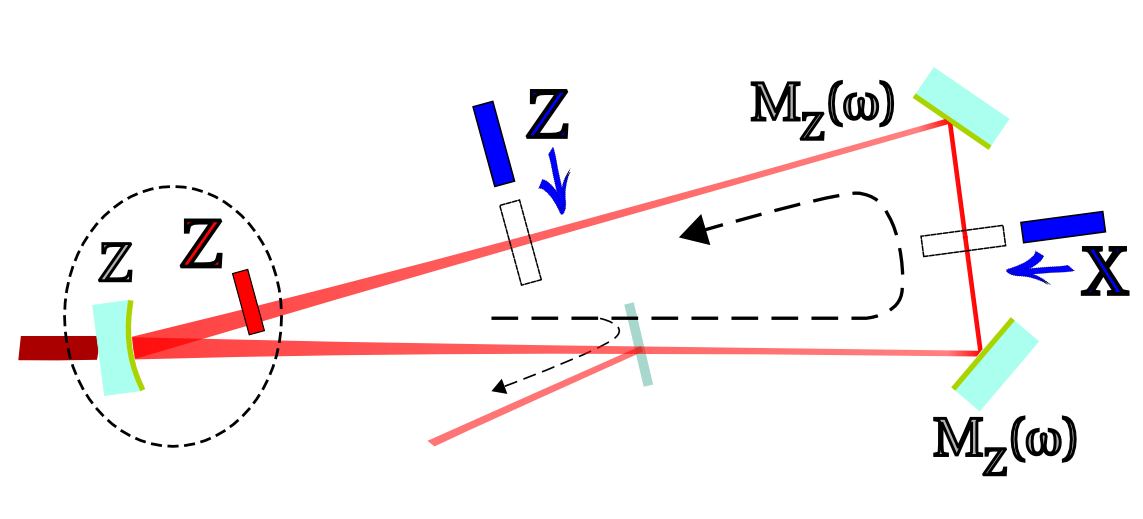}
   \caption{(Color online) Schematics of the triangular-ring cavity
   in the presence of BB Pauli group decoupling. An additional
   ${\mathbb Z}$ wave-plate is inserted in the long arm of the cavity,
   while a S--B with axis at 45$^\circ$ with respect to the cavity
   plane and delay equal to $\lambda/2$ is placed in the short arm
   of the cavity, acting therefore as $\mathbb{X}$.}
   \label{fig:BB_setup_Cavity_shortX_longZZ_2}
\end{figure}

The BB Pauli group dynamical decoupling is realized by adding two
control operations within the cavity: i) a second ${\mathbb Z}$
wave-plate in the long arm of the cavity in addition to the one
used for compensating the spherical mirror; ii) a S--B with axis
at 45$^\circ$ with respect to the cavity plane and delay equal to
$\lambda/2$ in the short arm of the cavity, acting therefore as
$\mathbb{X}$ (see Fig.~\ref{fig:BB_setup_Cavity_shortX_longZZ_2}).
The two controls implement every two cavity round-trips the full
Pauli-group decoupling for a polarization
qubit~\cite{Viola99,Massar2007}, and the overall transformation
for a double round-trip is given by
\begin{eqnarray}
   \mathbb{U}[\phi(\omega)]
   &=&\left[
            \mathbb{Z}\mathbb{M_Z}\mathbb{X}\mathbb{M_Z}
            \right]\cdot
            \left[
            \mathbb{Z}\mathbb{M_Z}\mathbb{X}\mathbb{M_Z}
            \right]
        \nonumber\\
   &=&[\mathbb{Z}\mathbb{X}]\, [\mathbb{Z}\mathbb{X}] = (i\mathbb{Y})^2 = -\mathbb{I}\,,
   \label{eq:CavityWithX_2}
\end{eqnarray}
where we have omitted the frequency dependence of $\mathbb{M_Z}(\omega)$.
\begin{figure}[!ht]
   \centering
   \includegraphics[width=.475\textwidth]{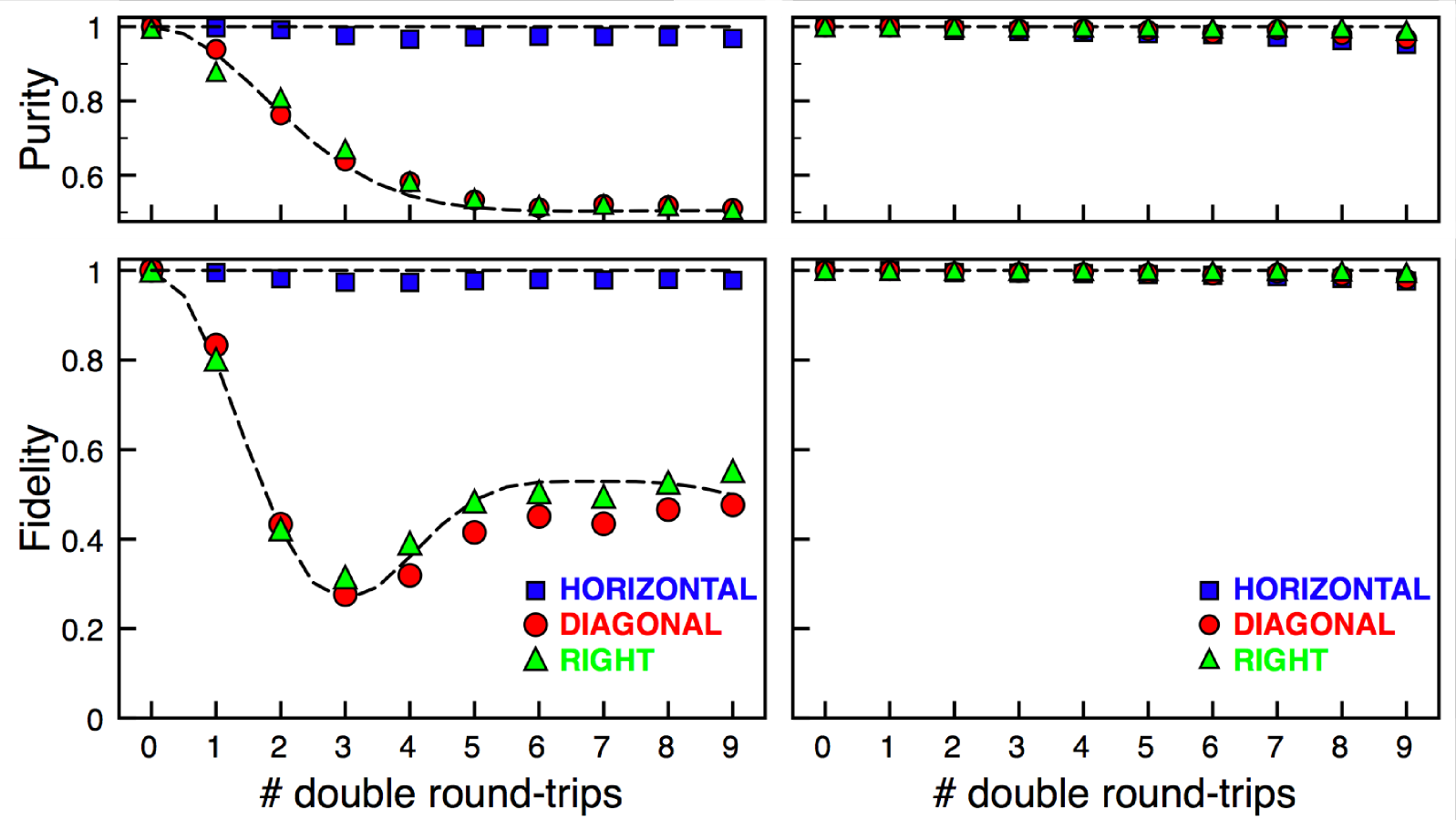}
   \caption{(Color online) Purity and fidelity as a function of the
   number of double round trips for three different input polarization
   states: left) cavity with  $\mathbb{Z}$ for compensating the mirror
   transformation; right) cavity with a $\mathbb{X}$ in the short arm,
   and a second $\mathbb{Z}$, as shown in Fig.~\ref{fig:BB_Cavity_X},
   for implementing the Pauli-group decoupling on the polarization-qubits.
   The dashed lines represent numerical simulation with the same parameters
   of the previous figures. (Adapted from Ref.~\cite{Damodarakurup2009}).}
   \label{fig:BB_PurFid_GeneralPauli}
\end{figure}
Therefore the polarization transformation is proportional to the
identity operator, thus implying a \textit{perfect} preservation
of every input polarization state, i.e., a complete suppression of
decoherence. This is experimentally verified for H, D and R input
polarizations in Fig.~\ref{fig:BB_PurFid_GeneralPauli}, where the
purity $\mathcal{P}$ and the fidelity $\mathcal{F}$ of the output
polarization state are plotted versus the number of double
round-trips. AS we have seen in the previous Section, when BB is
not performed (left panels of
Fig.~\ref{fig:BB_PurFid_GeneralPauli}), polarizations D and R
decay to the fully unpolarized state  ($\mathcal{P}= \mathcal{F} =
1/2$), while H, being an eigenstate of $\mathbb{Z}$, is unaffected
by decoherence. On the contrary, polarization decoherence is
completely suppressed when BB is applied (right panels of
Fig.~\ref{fig:BB_PurFid_GeneralPauli}).

\subsubsection{Carr--Purcell decoupling}
\begin{figure}[!ht]
   \centering
   \includegraphics[width=.375\textwidth]{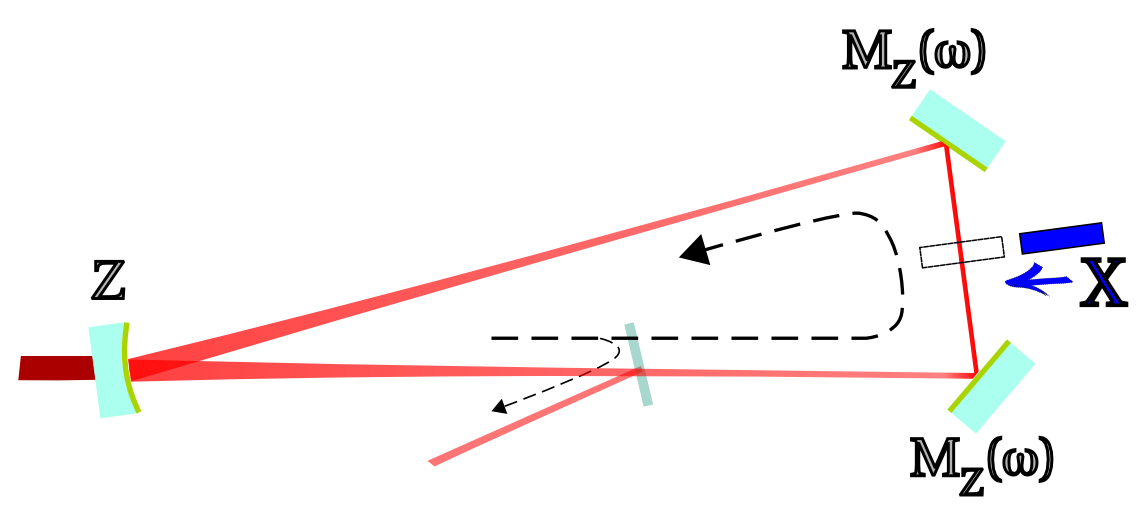}
   \caption{(Color online) Schematics of the triangular-ring cavity with a
   wave-plate inserted in the short arm of the cavity, which implements the
   $\mathbb{X}$ operation in order to realize the Carr-Purcell decoupling
   scheme for polarization-qubit.}
   \label{fig:BB_Cavity_X}
\end{figure}

\begin{figure}[!ht]
   \centering
   \includegraphics[width=.475\textwidth]{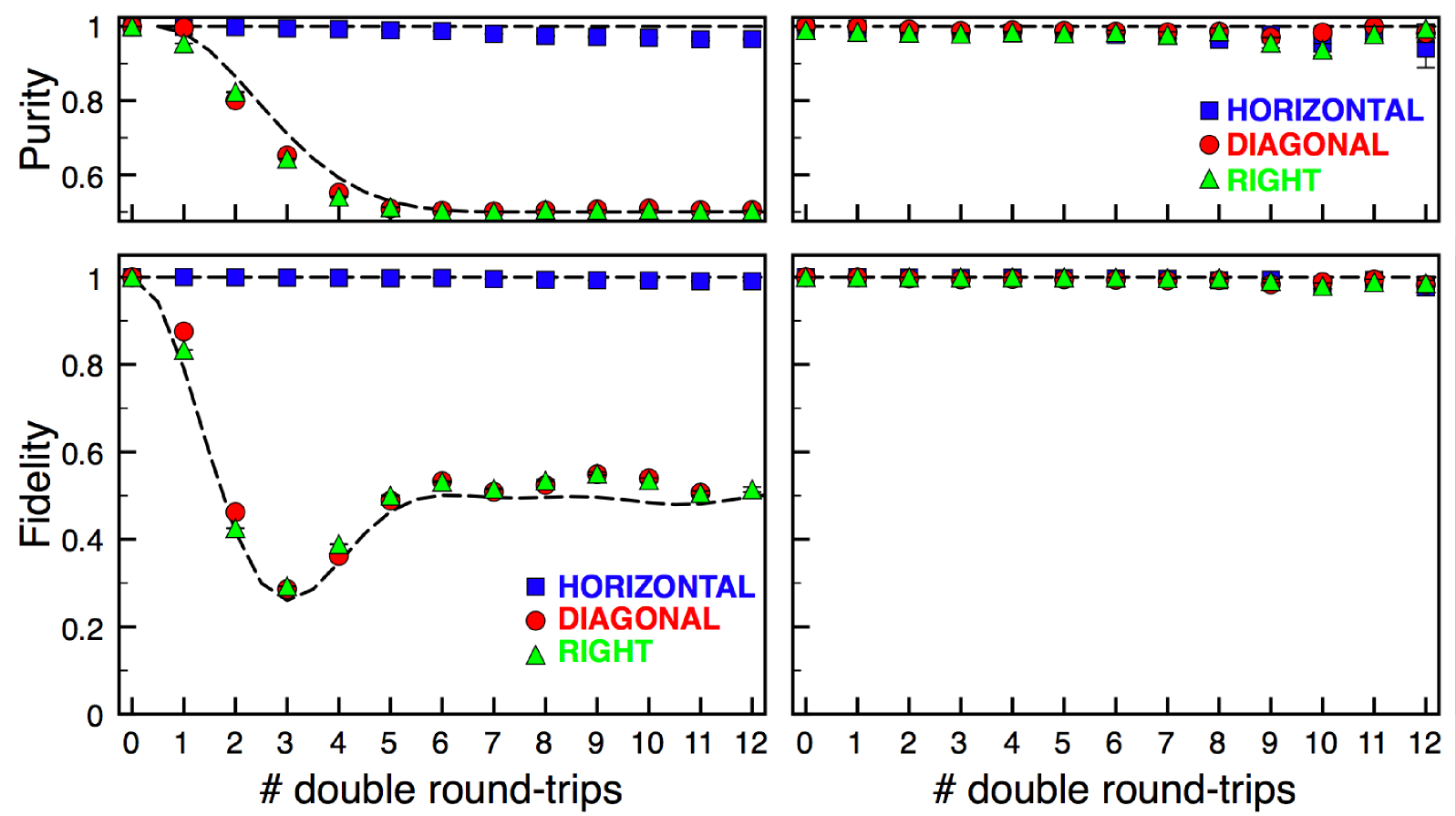}
   \caption{(Color online) Purity and fidelity versus the number of
   double round-trips for three different input polarization states: left)
   only cavity; right) cavity with a $\mathbb{X}$ in the short arm, as shown
   in Fig.~\ref{fig:BB_Cavity_X}, in order to realize the Carr-Purcell
   decoupling on polarization-qubits affected only by phase-noise along the
   $z$-axis. The dashed lines represent numerical simulation with the same
   parameters of the previous figures.}
   \label{fig:BB_PurityFidelityDouble_Compare_longZ_shortX}
\end{figure}

Polarization decoherence caused by a cavity round-trip acts along
a known axis of the Bloch sphere, the $z$-axis, and therefore even
the simplest Carr-Purcell decoupling scheme, based on the
decoupling group $\mathcal{G}=\{\mathbb{I},\mathbb{X}\}$, suffices
for suppressing decoherence. We have verified this fact by
inserting \textit{only} the $\mathbb{X}$ operation in the short
arm of the cavity, by properly adjusting the S--B (see
Fig.~\ref{fig:BB_Cavity_X}).
The Carr-Purcell decoupling scheme requires again a double
round-trip in the cavity, which gives the evolution
\begin{equation}\label{eq:CavityWithX}
   \mathbb{U}[\phi(\omega)]
   = \mathbb{Z}\mathbb{M_Z}
            \cdot\mathbb{X}\cdot
            \mathbb{M_Z}\mathbb{Z}\mathbb{M_Z}
            \cdot\mathbb{X}\cdot\mathbb{M_Z}
   = -\mathbb{I}\,.
\end{equation}
The polarization transformation is again proportional to the
identity operator, implying a perfect preservation of every input
polarization state. In fact, when the effective noise acts along a
\textit{known} axis, a complete suppression of decoherence is
already obtained by employing only \textit{one} Pauli operator,
the one realizing a spin-flip around an axis \textit{orthogonal}
to the axis of decoherence. This is very well shown in
Fig.~\ref{fig:BB_PurityFidelityDouble_Compare_longZ_shortX} where,
starting from input polarization states H, D and R, the purity
$\mathcal{P}$ and the fidelity $\mathcal{F}$ of the output
polarization state, with (right panels) and without (left panels)
Carr-Purcell decoupling are plotted versus the number of double
round-trips.

\subsection{Pauli-group decoupling for the most general polarization decoherence}
\begin{figure}[!ht]
   \centering
   \includegraphics[width=.375\textwidth]{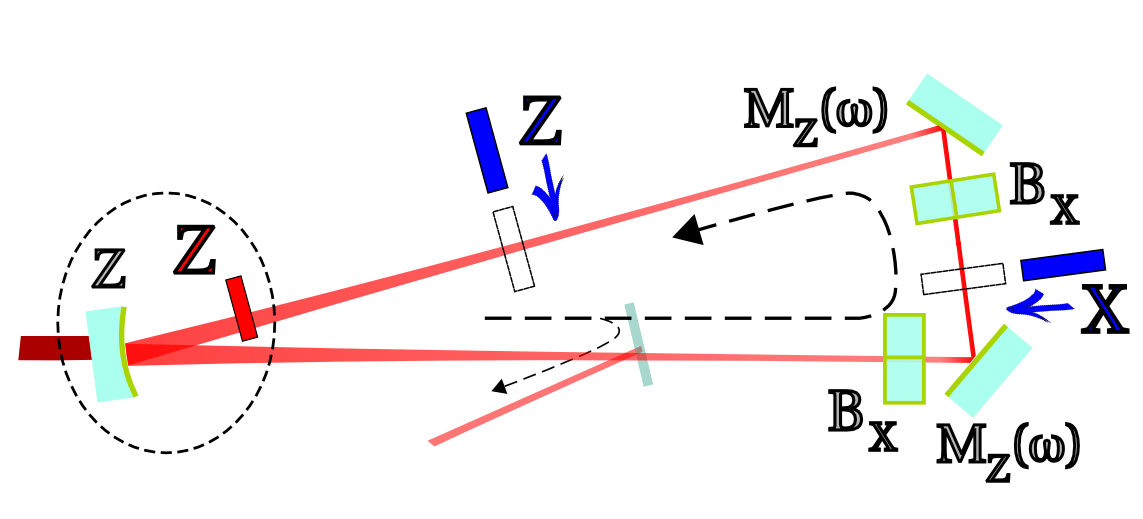}
   \caption{(Color online) Schematics of the experimental apparatus for generic noise.}
   \label{fig:BB_Cavity_GenericNoise}
\end{figure}

In the ring-cavity setup, the V and H polarization states are
eigenstates of the interaction Hamiltonian of the polarization
qubit with its effective environment and are therefore the
``pointer states'' unaffected by decoherence~\cite{Zurek2003}. In
the pointer states basis, decoherence affects only the
off-diagonal elements of the polarization density matrix. In the
most general case instead, the direction of the pointer state
basis in the Bloch--Poincar\'e sphere is \textit{unknown}. As a
consequence in a generic basis, decoherence acts \textit{both} on
diagonal and off-diagonal elements of the polarization density
matrix. To implement such a  generic noise model we have placed in
front of each plane mirror a S--B with axis at 45$^\circ$ with
respect to the cavity plane (see
Fig.~\ref{fig:BB_Cavity_GenericNoise}). The action of the S--B on
the polarization state is described by
\begin{equation}
    \mathbb{B_X}[\theta] = \exp[-{\rm i}\theta \mathbb{X}/2]\,,
\end{equation}
where $\mathbb{X}=|H\rangle \langle V|+|V\rangle \langle H|$, and
$\theta$ is the noise delay-phase.
The S-B together with a plane mirror are described by the operator
\begin{equation}
    \mathbb{N}(\omega,\theta) = \mathbb{M_Z}(\omega)\mathbb{B_X}(\theta)\,.
\end{equation}
The transformation of the polarization state after two round-trips
in the presence of the S--Bs is therefore given by
\begin{equation}
    \mathbb{U}_{fe}[\phi(\omega),\theta] =
        [\mathbb{N}(\omega,\theta)\mathbb{N}(\omega,\theta)]\cdot
        [\mathbb{N}(\omega,\theta)\mathbb{N}(\omega,\theta)]\,.
\end{equation}
The free evolution ($fe$) operator
$\mathbb{U}_{fe}[\phi(\omega),\theta]$ can be rewritten as
\begin{equation}
    \mathbb{U}_{fe}[\phi(\omega),\theta] =
        \exp\left[-i \alpha_{fe}(\omega,\theta)\,\vec{s}_{fe}(\omega,\theta)\cdot \vec{\sigma}\right]\,,
\end{equation}
which describes a rotation in the Bloch--Poincar\'e sphere of an
angle $2\,\alpha_{fe}(\omega,\theta)$ around the direction
individuated by $\vec{s}_{fe}(\omega,\theta)$. Therefore by
varying $\theta$ and the bandwidth of radiation spectrum, i.e. the
distribution of $\phi$, one implements the generic polarization
decoherence. The rotation angle is given by the implicit
expression
\begin{equation}\label{eq:alphano}
        \sin [\alpha_{fe}(\omega,\theta)/2]=\sin [\phi(\omega)/2]\cos(\theta/2),
\end{equation}
and
\begin{eqnarray}
    \vec{s}_{fe}(\omega,\theta)= \frac{1}{2\sin\alpha_{fe}(\omega,\theta)}
    \{\!\!\!&&\!\!\!\sin\theta[\cos\phi(\omega)-1]\,, \nonumber\\
    \!\!\!&&\!\!\!\sin\theta\sin\phi(\omega)\,,\nonumber\\
    \!\!\!&&\!\!\!(1+\cos\theta)\sin\phi(\omega)\,\}.
\end{eqnarray}
Pauli-group decoupling is again realized every two-round trips by
adding the BB operations $\mathbb{X}$ (implemented by adding to
the noise phase-delay $\theta$ of the S--B present in the short
arm of the cavity, a further delay of $\lambda/2$) and
$\mathbb{Z}$ in the cavity~\cite{Viola99,Massar2007} as shown in
Fig.~\ref{fig:BB_Cavity_GenericNoise}.
The overall transformation after the two round-trips for a given
frequency component is now given by
\begin{equation}
   \mathbb{U}_{bb}[\phi(\omega),\theta] =
    [\mathbb{Z}\mathbb{N}(\omega,\theta)\mathbb{X}\mathbb{N}(\omega,\theta)]
    \cdot
    [\mathbb{Z}\mathbb{N}(\omega,\theta)\mathbb{X}\mathbb{N}(\omega,\theta)]\,,
\end{equation}
which can be rewritten, as in the free evolution case, as a
rotation in the Bloch--Poincar\'e sphere,
\begin{equation}
   \mathbb{U}_{bb}[\phi(\omega),\theta] =
    \exp[-i\alpha_{bb}(\omega,\theta)\,\vec{s}_{bb}(\omega,\theta)\cdot \vec{\sigma}]\,,
\end{equation}
with
\begin{equation}\label{eq:implicitalph}
\cos \alpha_{bb}(\omega,\theta)=-(\sin\phi(\omega)\,\sin\theta)/2,
\end{equation}
and
\begin{eqnarray}
    \vec{s}_{bb}(\omega,\theta)=
    \!\!\!&&\!\!\!  \{-\sin\phi(\omega)\sin^2(\theta/2)\,, \nonumber\\
    \!\!\!&&\!\!\!  1-2\sin^2(\theta/2)\sin^2[\phi(\omega)/2]\,,  \nonumber\\
    \!\!\!&&\!\!\!  -\sin\theta\sin^2[\phi(\omega)/2]\}/[\sin\alpha_{bb}(\omega,\theta)].
\end{eqnarray}

\subsubsection{Suppressing decoherence of an input elliptical polarization state}
\begin{figure}[!ht]
   \centering
   \includegraphics[width=.475\textwidth]{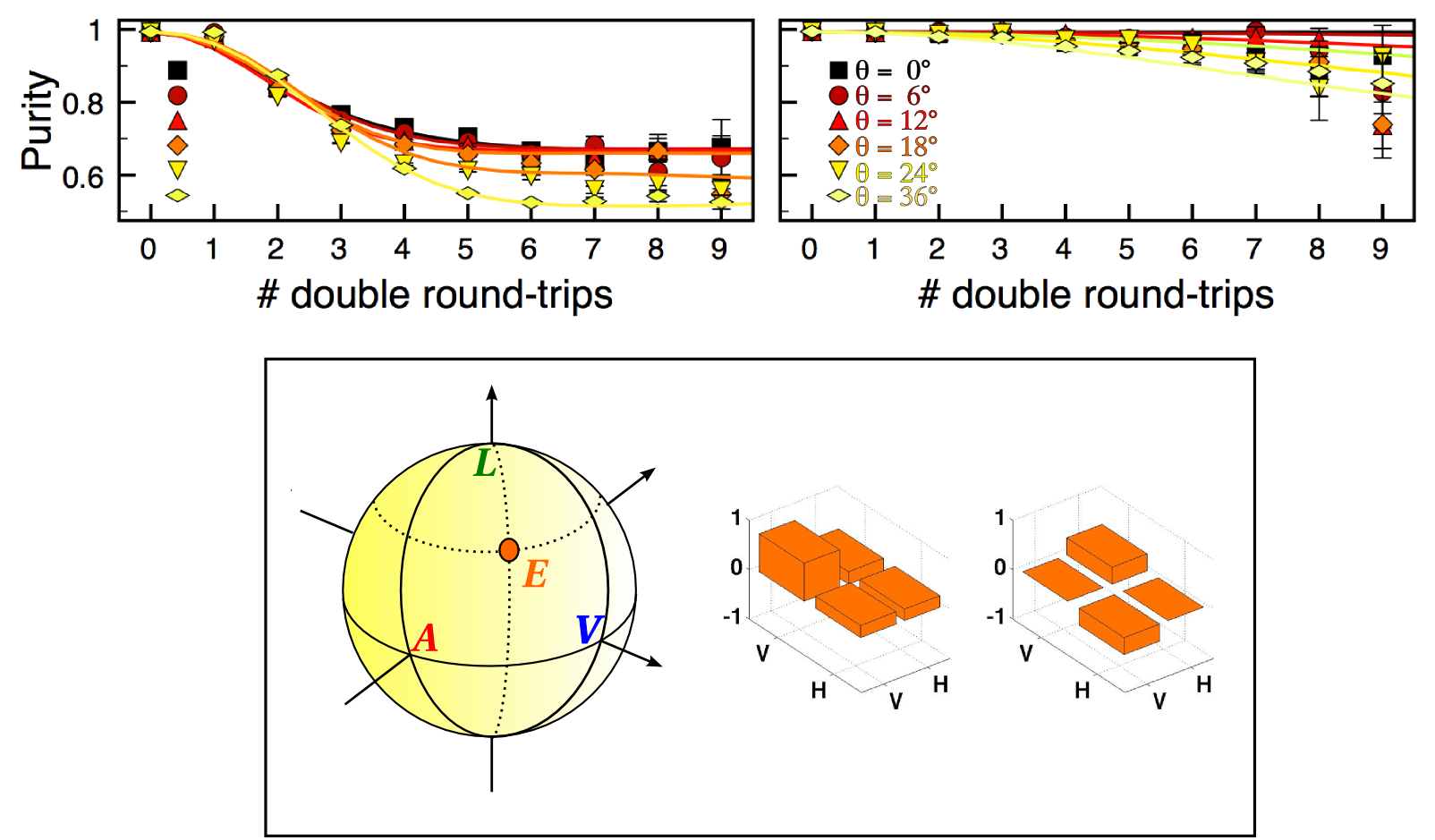}
   \caption{(Color online) Purity as a function of the number of double
   round trips for different values of the noise delay-phase $\theta$ for
   an elliptical polarization input state. The lines represent numerical
   simulation with the parameters of the previous figures. Inset: real and
   imaginary parts of the density matrix of the elliptical polarization input
   state, also shown on the Bloch--Poincar\'e sphere as the point labeled $E$.
   (Adapted from Ref.~\cite{DamodarakurupSPIE2009}).}
   \label{fig:BB_GenericNoise_3.0_HALF}
\end{figure}

We have performed a first test of the ability of Pauli group
decoupling for controlling decoherence under generic noise, by
injecting in the cavity an elliptical polarization state,
corresponding to an equal-weighted superposition of V, A
(\textit{anti-diagonal}) and R polarizations (see the
experimentally reconstructed density matrix and Bloch-Poincar\'e
sphere representation in Fig.~\ref{fig:BB_GenericNoise_3.0_HALF}).
In case of no BB control, the resulting fast polarization decay of
the purity, as a function of the number of double round-trips, is
shown in the left panel of
Fig.~\ref{fig:BB_GenericNoise_3.0_HALF}, for different noise
delay-phases $\theta$. The recovery of purity after the
application of BB is evident from the right panel of the
Fig.~\ref{fig:BB_GenericNoise_3.0_HALF}.

\subsubsection{Purity and fidelity averaged over the Bloch--Poincar\'e sphere}
\begin{figure}[!ht]
   \centering
   \includegraphics[width=.475\textwidth]{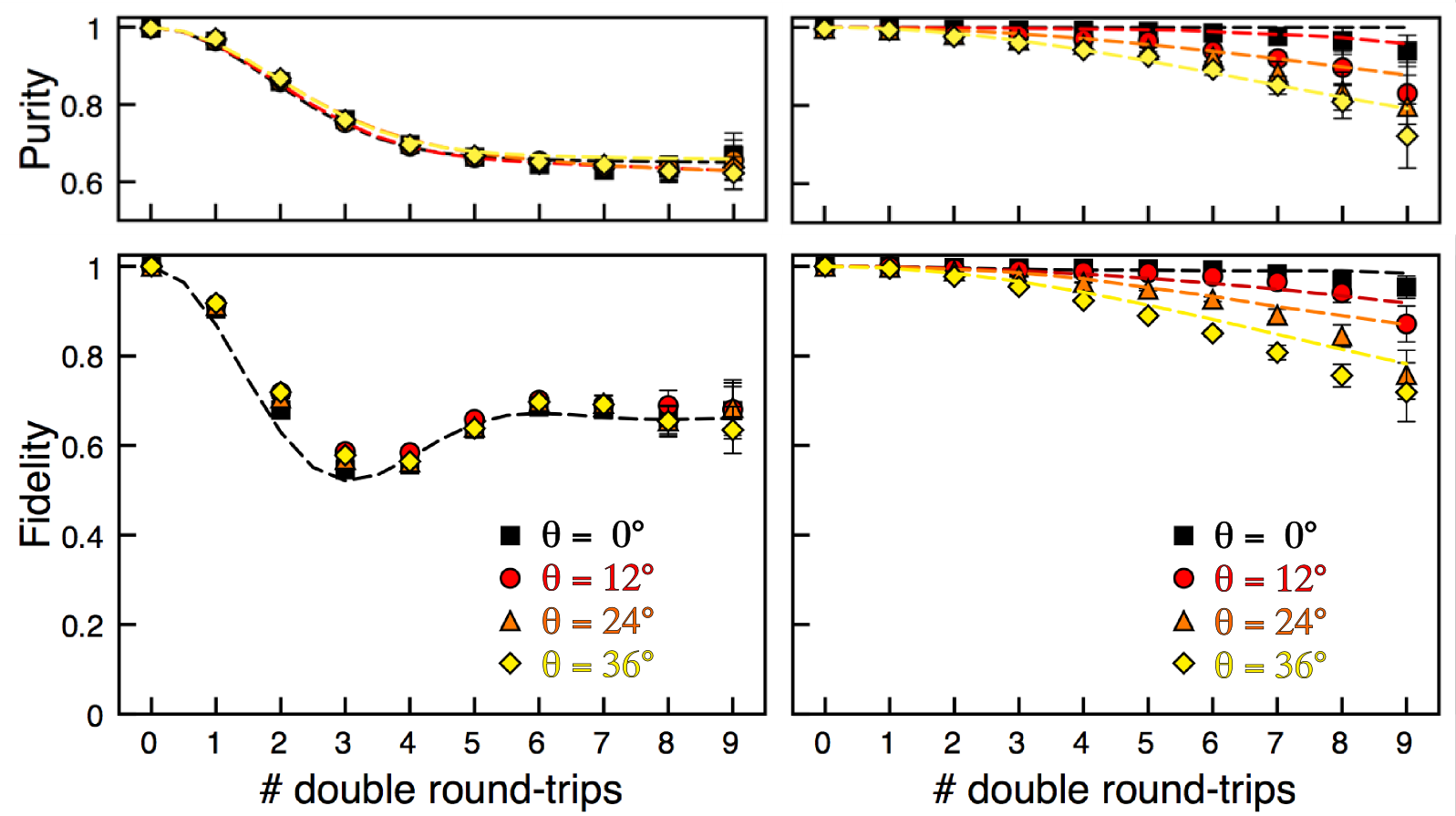}
   \caption{(Color online) Average purity and fidelity as a function of the number of double round trips for different values of the generic noise phase-delay parametert, $\theta$: left) without bang-bang control: right) when applied the bang-bang control. The lines represent numerical simulation with the parameters of the previous figures. (Adapted from Ref.~\cite{Damodarakurup2009}).}
   \label{fig:BB_PurFid_Average_Compare}
\end{figure}

The experiment with the input elliptical polarization state
already shows the effectiveness of the BB Pauli group decoupling
for suppressing decoherence. However, in order to be useful for
any quantum information application, the scheme must work for
every input polarization. Therefore, in order to better highlight
the usefulness of BB decoupling to combat generic
birefringence-induced decoherence, we have studied the evolution
of the purity and the fidelity, averaging the input state over the
whole Bloch--Poincar\'e sphere, both with (right panels of
Fig.~\ref{fig:BB_PurFid_Average_Compare}) and without BB
decoupling (left panels of
Fig.~\ref{fig:BB_PurFid_Average_Compare}).
Each curve corresponds to a different orientation of the noise
delay-phase $\theta$ of the S-B. BB again inhibits decoherence
because, for each orientation of the decoherence-axis, both the
average purity and the average fidelity in the presence of BB are
significantly higher than the corresponding value without BB. The
curve for $\theta =0$ reproduces the almost perfect preservation
of the previous experiment of Sec.~IVA. Instead polarization
protection progressively worsens for increasing values of
$\theta$. When $\theta$ is varied, the properties of the effective
environment of the polarization qubit associated with the
effective birefringence caused by the mirrors changes in a
nontrivial way, but a simple analytical explanation of the result
can be given when the pulses are not too broad in frequency.

\subsubsection{Analytical explanation of the results}

An approximate analytical description of the evolution of the
polarization state as it circles within the cavity can be given
under the assumption that the pulse is not too broad in frequency,
$\sigma_{\phi} \ll \pi$, which is well verified in the present
experiment. In this case, $\alpha_{j}$ and $\vec{s}_{j}$
($j=fe,bb$) do not vary appreciably over the range of relevant
phase shifts $\phi$ and one can approximate $\vec{s}_j(\phi)$ with
its value at the pulse center $\vec{s}_j(\phi_0)$, while
$\alpha_j(\phi)$ can be approximated by its second-order expansion
around $\phi_0$,
\begin{equation}
    \alpha(\phi)\simeq \alpha_0+
                     \dot{\alpha}_0 (\phi-\phi_0)+
                     \frac{1}{2}\ddot{\alpha}_0 (\phi-\phi_0)^2\,.
\end{equation}
After lengthy but straightforward calculations, one gets for the
output Bloch vector $\vec{P}_{out}$, which determines completely
the density matrix as expressed in Eq.~\eqref{eq:rhopolarization},
\begin{equation}
    \vec{P}_{out} = V^{(n)} \vec{P}_{in}\,.
\end{equation}
The matrix $V^{(n)}$ is given by
\begin{equation}\label{eq:purity3}
    V^{(n)}_{ij} = {\cal D}_{n}O^{(n)}_{ij}+\left(1-{\cal D}_{n}\right)s(\phi_0)_i s(\phi_0)_j\,
\end{equation}
($i,j=x,y,z$), where the ``decoherence factor'' ${\cal D}_{n}$ is given by
\begin{equation}\label{eq:purityD}
    {\cal D}_{n} = \left[1+n^2 \ddot{\alpha}_0^2 \sigma_{\phi}^4\right]^{-1/4}
                \exp\left[-\frac{n^2 \dot{\alpha}_0^2 \sigma_{\phi}^2}
                {1+n^2 \ddot{\alpha}_0^2 \sigma_{\phi}^4}\right],
\end{equation}
and the matrix
\begin{eqnarray}
   O^{(n)}_{ij}\!\!\!&&=\delta_{ij}\cos 2n\gamma_n
    \nonumber\\
   &&+s(\phi_0)_i s(\phi_0)_j\left(1-\cos 2n\gamma_n\right)\nonumber\\
   &&-\varepsilon_{ijk}s(\phi_0)_k \sin2n\gamma_n \label{eq:purityO}
\end{eqnarray}
is the orthogonal matrix describing the rotation around the $\vec{s}(\phi_0)$-axis of an angle $2 \gamma_n$, with
\begin{equation}\label{eq:purityGam}
    \gamma_n=\alpha_0+\frac{1}{2}\frac{n^2 \ddot{\alpha}_0 \dot{\alpha}_0^2 \sigma_{\phi}^2}
    {1+n^2 \ddot{\alpha}_0^2 \sigma_{\phi}^4}+\frac{1}{4n}\arctan( n \ddot{\alpha}_0 \sigma_{\phi}^2)\,.
\end{equation}
The output purity and fidelity are respectively given by
\begin{eqnarray}
    \mathcal{P}_n
    \!\!\!&=&\!\!\! \frac{1}{2}\left(1+\vec{P}_{out}\cdot \vec{P}_{out}\right)\nonumber \\
    \!\!\!&=&\!\!\! \frac{1}{2}\left\{1+2{\cal D}_n (1-{\cal D}_n)\left[\vec{P}_{in}\cdot \vec{s}(\phi_0)\right]
                \left[\vec{P}(\gamma_n)\cdot \vec{s}(\phi_0)\right]\right. \nonumber \\
    \!\!\!&&\!\!\! \left.+{\cal D}_n^2 +(1-{\cal D}_n)^2\left[\vec{P}_{in}\cdot \vec{s}(\phi_0)\right]^2\right\}\,,
\end{eqnarray}
and
\begin{eqnarray}
    {\mathcal F}_n
    \!\!\!&=&\!\!\! \frac{1}{2}\left(1+\vec{P}_{in}\cdot \vec{P}_{out}\right)\\
    \!\!\!&=&\!\!\! \frac{1}{2}\left\{1+{\cal D}_n \vec{P}_{in} \cdot \vec{P}(\gamma_n)
                +(1-{\cal D}_n)\left[\vec{P}_{in}\cdot \vec{s}(\phi_0)\right]^2\right\} \,,\nonumber
\end{eqnarray}
with $\vec{P}(\gamma_n)=O^{(n)} \vec{P}_{in}$. These expressions
are general and apply both to the BB case as well to the one
without BB. The two cases differ only in the explicit expressions
of $\vec{s}(\phi_0)$, $\alpha_0$, $\dot{\alpha}_0$, and
$\ddot{\alpha}_0$.
Since $\lim_{n \to \infty} {\cal D}_n=0$, fidelity and purity tend asymptotically to the same limit
\begin{equation}
    \mathcal{P}_{\infty}= \mathcal{F}_{\infty}=\frac{1}{2}\{1+[\vec{P}_{in}\cdot \vec{s}(\phi_0)]^2\}\,.
\end{equation}
This asymptotic situation corresponds to the existence of a
pointer basis~\cite{Zurek2003} (analogous to the principal states
of polarization in fibers \cite{Poole88}), formed by the two
states with Bloch vector equal to $\pm \vec{s}(\phi_0)$, which are
unaffected by decoherence. All the other polarization states
instead decohere, with maximum decoherence for the states on the
plane orthogonal to $\vec{s}(\phi_0)$. As a consequence, when
averaged over the initial state, purity and fidelity tends to
$2/3$. In the $n \to\infty$ regime, BB and no-BB case differ only
in the direction $\vec{s}(\phi_0)$ of the pointer basis and
therefore BB is not advantageous with respect to the no-BB case in
this regime.

However, the interesting regime of our experiment is the one
corresponding to a small round-trip number $n$. In that regime our
ring-cavity well mimics a portion of a one-way quantum
communication channel with constant dispersive properties, like a
small portion of an optical fiber. For small $n$ one has
\begin{eqnarray}
    1-\mathcal{F} \!\!\!& \simeq &\!\!\! (1-\mathcal{P})/2\nonumber\\
                  \!\!\!& \simeq &\!\!\! n^2 \left[\dot{\alpha}_j^0 \sigma_{\phi}\right]^2
                  \{1-[\vec{P}_{in}\cdot \vec{s}_j(\phi_0)]^2\}/2\,,
\end{eqnarray}
with $j=\{fe,bb\}$, showing that the smaller $\dot{\alpha}^0$ the
better the decoherence suppression. BB Pauli-group decoupling acts
just by decreasing $ |\dot{\alpha}^0 |$: by using
Eqs.~(\ref{eq:alphano})-(\ref{eq:implicitalph}) and the fact that
$\phi_0$ is quite small in our experiment, one gets
$\dot{\alpha}_{fe}^0 \simeq \cos(\theta/2)$,
$\dot{\alpha}_{bb}^0 \simeq \sin(\theta/2)\cos(\theta/2)$,
showing that it is always $\dot{\alpha}_{fe}^0 >
\dot{\alpha}_{bb}^0$ and therefore that BB better preserves the
polarization qubit for any orientation of the decoherence axis,
confirming its applicability to otpical fibers polarization
control.
These expressions also explain the perfect preservation of the
polarization qubit of Fig.~\ref{fig:BB_PurFid_GeneralPauli}: the
latter refers to $\theta =0$, implying $\dot{\alpha}_{fe}^0 =1$
and $\dot{\alpha}_{bb}^0=0$.

\section{Conclusions}

We have presented the details of the experiment of
Ref.~\cite{Damodarakurup2009}, which showed how BB dynamical
decoupling can be efficiently implemented in the optical domain in
order to suppress the polarization decoherence caused by the
propagation of light pulses within birefringent media. In such
media, the optical properties depend upon both frequency and
polarization and the frequency degree of freedom acts as an
effective environment on the photon polarization. With photons BB
dynamical decoupling can be realized by placing appropriate
optical elements (i.e., wave-plates) implementing suitable unitary
operations along the photon path. In the present
proof-of-principle demonstration, a one-way communication channel
is mimicked by a ring-cavity, in which an effective birefringence
is introduced by two mirrors at $45$ degrees whose reflectivity is
responsible for the decay of the density matrix off-diagonal
elements, in the basis formed by H and V polarizations. In a first
experiment, BB dynamical decoupling was shown to be able to
perfectly suppress this kind of decoherence. In a second
experiment, we modified the setup by placing two Soleil--Babinet
compensators in front of the mirrors. This allowed us to simulate
the most general model of decoherence, for which the pointer basis
is unknown. Also in this case the Pauli group BB decoupling was
able to significantly reduce decoherence, as predicted
by~\cite{Viola1999}.

Several elements suggest that our experiment is significant for
the potential extension of BB decoupling to the dynamics of
polarization pulses propagating in optical fibers. First, the
interaction Hamiltonian governing the dynamics in the ring cavity
is strictly related to the one causing PMD in optical
fibers~\cite{Massar2007}. Second, the one-way nature of the ring
cavity well mimics a fiber-based communication channel. Third, the
optical elements introducing decoherence in the ring-cavity
doubtless realize a worst-case version of the fiber dynamics,
because the errors they introduce accumulate systematically while
in a fiber they are distributed at random.

We plane to experimentally investigate the optimal length for
placing the BB operations along an SM fiber, obtained from a
tradeoff between the two opposite requests of minimizing the
attenuation of the fiber and maximizing the decoherence
suppression.

\section{Acknowledgments}
We acknowledge the funding of the EC project FP6-IP-QAP. M.L. is
supported by the $5\tcperthousand$~grant C.F.~81001910439.



\end{document}